\documentclass[aip,reprint]{revtex4-1}
\usepackage{graphicx}
\usepackage{epsfig}
\usepackage{color}
\usepackage{epstopdf}
\usepackage{amsmath}
\usepackage[varg]{txfonts}

\newcommand{\grl}{    {Geophys. Res. Lett. }}

\newcommand{\planss}{    {Planet. Space. Sci.}}
\newcommand{\aap}{    { Astronomy and Astrophysics }}

\newcommand{\apj}{ {Astrophys. J. }}

\newcommand{\prl}{ {Phys. Rev. Lett. }}
\newcommand{\apss}{ {Astrophys. \& Space Sci. }}
\newcommand{\pre}{ {Phys. Rev. E }}
\newcommand{\physscr}{ {Physica Scripta}}

\begin{document}


\title{The separation of ions and fluxes in nonlinear ion-acoustic waves} 



\author{A.~E.~Dubinov}
\affiliation{Sarov Institute of Physics and Technology (SarFTI), National Research Nuclear University
	\lq\lq MEPhI\rq\rq, Sarov, 607188, Russia}

\author{I.~N.~Kitayev}
\affiliation{Sarov Institute of Physics and Technology (SarFTI), National Research Nuclear University
\lq\lq MEPhI\rq\rq, Sarov, 607188, Russia}

\author{D.~Y.~Kolotkov}
\altaffiliation[Corresponding author: ]{d.kolotkov.1@warwick.ac.uk}
\affiliation{Centre for Fusion, Space and Astrophysics, Physics Department, University of Warwick, Coventry, CV4 7AL, UK}
\affiliation{Institute of Solar-Terrestrial Physics SB RAS, Irkutsk, 664033, Russia}

\date{\today}

\begin{abstract}
{The multi-species plasma of natural or laboratory origin is often considered to host nonlinear ion-acoustic waves.}
We present calculations of ion fluxes induced by nonlinear ion-acoustic waves in a plasma consisting of multiple ion populations, electrons, and dust. The following plasma models are considered: an electron-ion plasma with cold ions, a bi-ion plasma with two types of warm positively charged ions, and a dusty bi-ion plasma. It is found that in the electron-ion plasma, the wave-induced ion flux is directed oppositely to the phase speed of the nonlinear ion-acoustic wave. In the bi-ion plasma, there are two modes of ion-acoustic waves which are fast and slow waves. In the nonlinear fast ion-acoustic wave, the fluxes of both types of ions are found to be co-directed and drift against the wave. In a slow wave, the nonlinear fluxes of ions are directed in opposite directions. This result demonstrates the possibility to use these nonlinear wave-induced ion fluxes for effective separation of ions in the plasma. In a dusty bi-ion plasma, the ion separation process can be intensified by a super-nonlinear regime of slow ion-acoustic waves.

\end{abstract}

\maketitle

\section{Introduction}

Periodic stationary cnoidal ion-acoustic waves in a collisionless electron-ion plasma (\textit{ei}-plasma) with cold ions are well known to carry a non-zero ion flux averaged over the wave period \cite{1979JPSJ...46.1907K}. In the follow up works~\onlinecite{2007PhPl...14b2106T, 2009PlPhR..35..651P, 2012PhPl...19j3702J, 2016Ap&SS.361..292U, 2019Prama..92...86K}, cnoidal ion-acoustic waves were considered for other models of collisionless plasma, and average ion fluxes associated with the wave were calculated: in an \textit{ei}-plasma with warm ions \cite{2007PhPl...14b2106T}, and in a dusty \textit{eid}-plasma with warm and cold ions\cite{2009PlPhR..35..651P, 2012PhPl...19j3702J, 2016Ap&SS.361..292U, 2019Prama..92...86K}. The presence of ion fluxes in electrostatic waves implies not only the transport of ions along the direction of the wave propagation, but also the induction of the ion flux-caused magnetic field.

More specifically, Ref.~\onlinecite{2007PhPl...14b2106T} demonstrated that a nonlinear wave of a relatively low amplitude carries a positive average ion flux (i.e. an ion flux co-directed with the wave phase speed $V$), while a large-amplitude wave leads to a negative ion flux directed against the wave propagation.  A similar dependence of the sign of the nonlinear wave-induced ion flux on the wave amplitude was obtained in Ref.~\onlinecite{2012PhPl...19j3702J}, while in Ref.~\onlinecite{2009PlPhR..35..651P} the flux was found to be always positive. Similarly to the discussed ion fluxes in nonlinear electrostatic waves, nonlinear Alfv\'en waves are also known to cause field-aligned perturbations of the plasma density and generate parallel plasma flows also known as the Alfv\'enic wind (see e.g. Refs.~\onlinecite{2011A&A...526A..80V, 2012A&A...544A.127V, 2012A&A...544A.127V, 2017ApJ...840...64S}, for the consideration of this phenomenon in the field-aligned plasma structures of the Sun's corona). Likewise, numerical simulations of nonlinear Alfv\'en and low-frequency acoustic waves in a partially ionised two-fluid plasma of the lower solar atmosphere demonstrated that those waves can be responsible for plasma outflows into the corona and origin of the solar wind (see e.g. Refs.~\onlinecite{2019ApJ...878...81K, 2020A&A...639A..45K}, and references therein).

In works~\onlinecite{1979JPSJ...46.1907K,2007PhPl...14b2106T, 2009PlPhR..35..651P, 2012PhPl...19j3702J, 2016Ap&SS.361..292U, 2019Prama..92...86K}, plane cnoidal ion-acoustic waves were considered as solutions of the corresponding evolutionary equations of the Korteweg-De Vries (KdV) type. It is known that such equations are derived from the equations of ion dynamics and analysed using the reductive perturbation method and expansion in a small parameter (see e.g. Ref.~\onlinecite{1987AuJPh..40..593M, 1988AuJPh..41....1M}), whereas the wave-induced fluxes are usually described by the second and higher-order terms. Thus, motivated by the recent Parker Solar Probe observations of nonlinear ion-acoustic waves around boundaries of the heliospheric magnetic switchbacks\cite{2021ApJ...911...89M}, in this work we determine exact values of the ion fluxes averaged over the wave period in nonlinear ion-acoustic waves of arbitrary amplitude using the Sagdeev pseudopotential method\cite{1966RvPP....4...23S, 1979RvMP...51...11S}, which allows for an accurate analysis of the ion dynamics without applying the approximate expansions. 

\begin{figure*}
	\begin{center}
		\includegraphics[width=\linewidth]{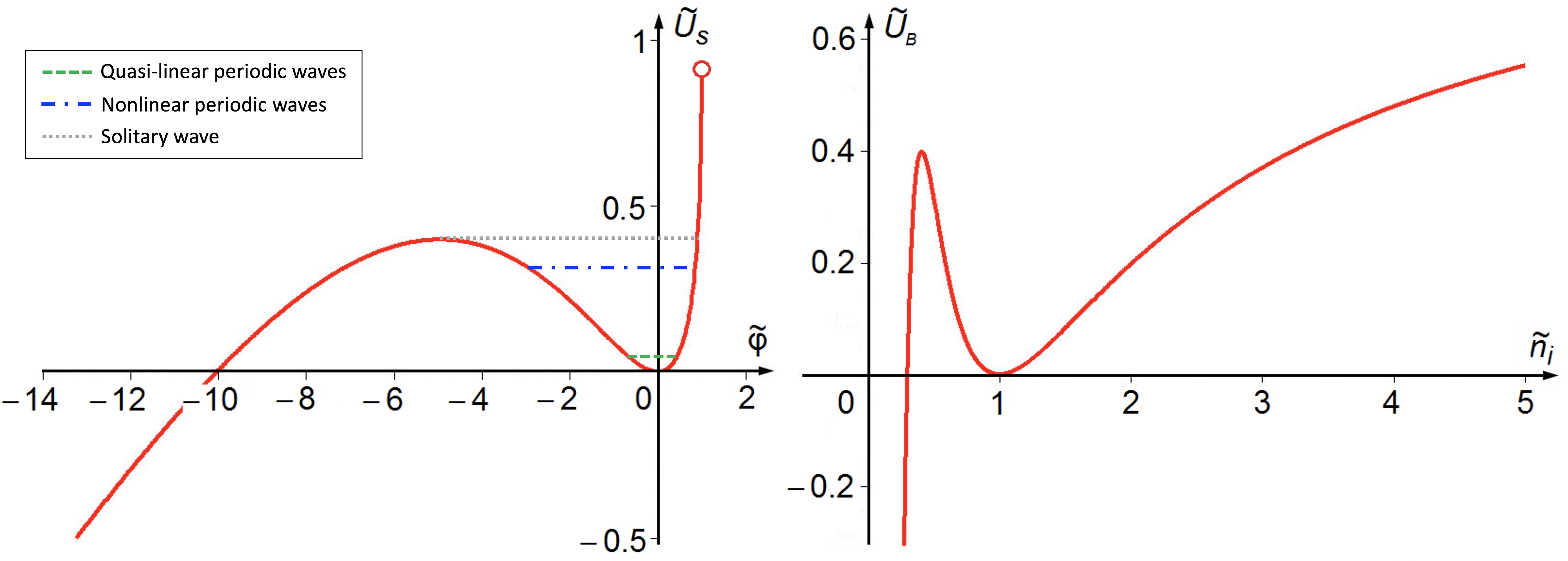}
	\end{center}
	\caption{The Sagdeev pseudopotential $U_S(\varphi)$ given by Eq.~(\ref{eq:ei_sagdeev}) and pseudopotential $U_B(n_i)$ obtained from Eqs.~(\ref{eq:ei_sagdeev}) and (\ref{eq:ei_n(phi)}) for the \textit{ei}-plasma model, normalised as $\tilde{U}_S=U_S/(2\pi n_{0i}m_iV^2)$, $\tilde{\varphi}=-2e\varphi/(m_iV^2)$, $\tilde{U}_B=U_B/(2\pi n_{0i}m_iV^2)$, and $\tilde{n}_i=n_i/n_{0i}$.
		{The examples of the energy levels corresponding to small-amplitude quasi-linear and high-amplitude nonlinear periodic waves around the initial equilibrium $\varphi=0$ are shown in green and blue, respectively. The energy level of a solitary wave (not considered in this work) is shown in grey.}
	}
	\label{fig:2+3}
\end{figure*}

\begin{figure*}
	\begin{center}
		\includegraphics[width=0.51\linewidth]{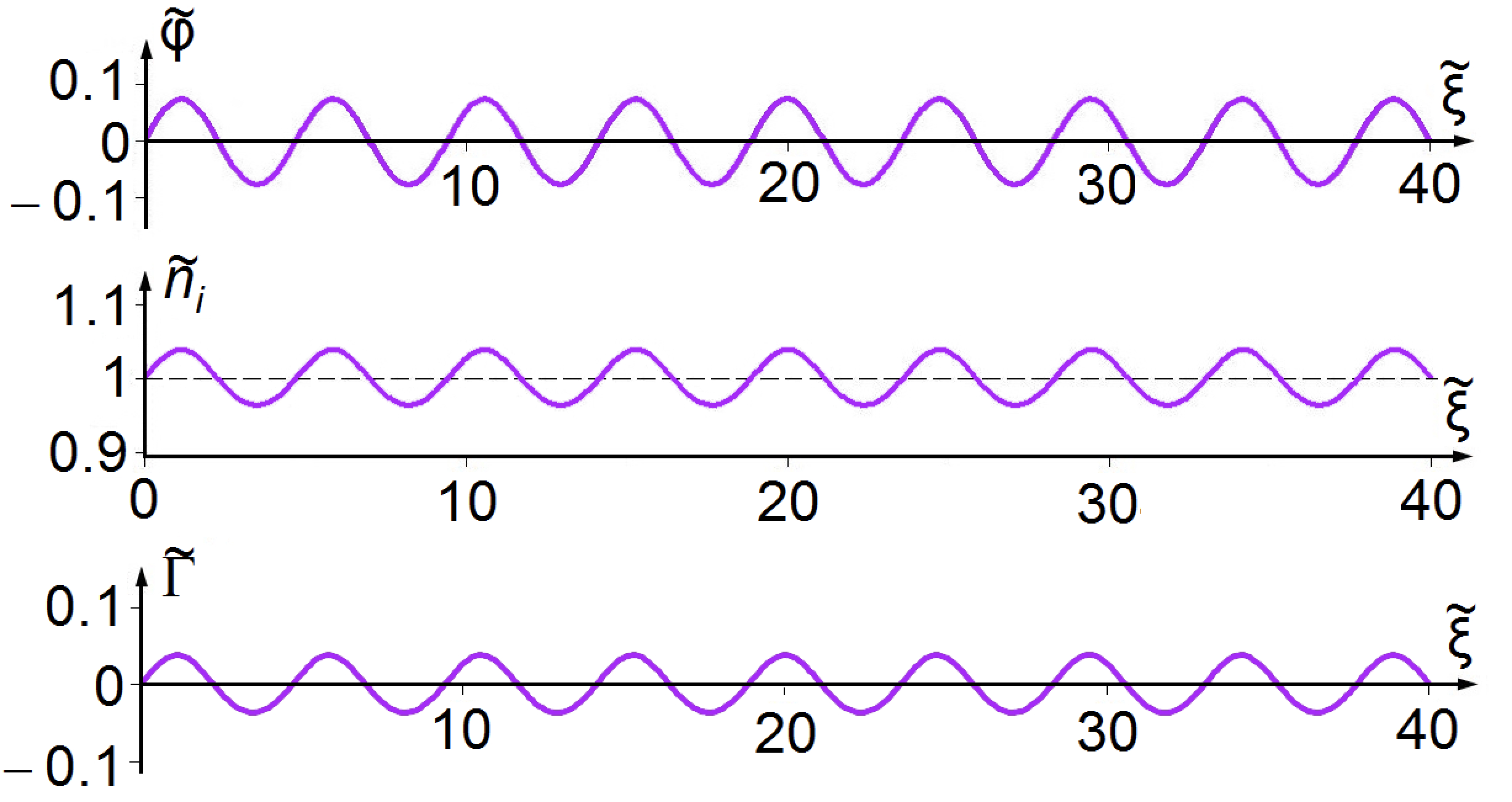}
		\includegraphics[width=0.47\linewidth]{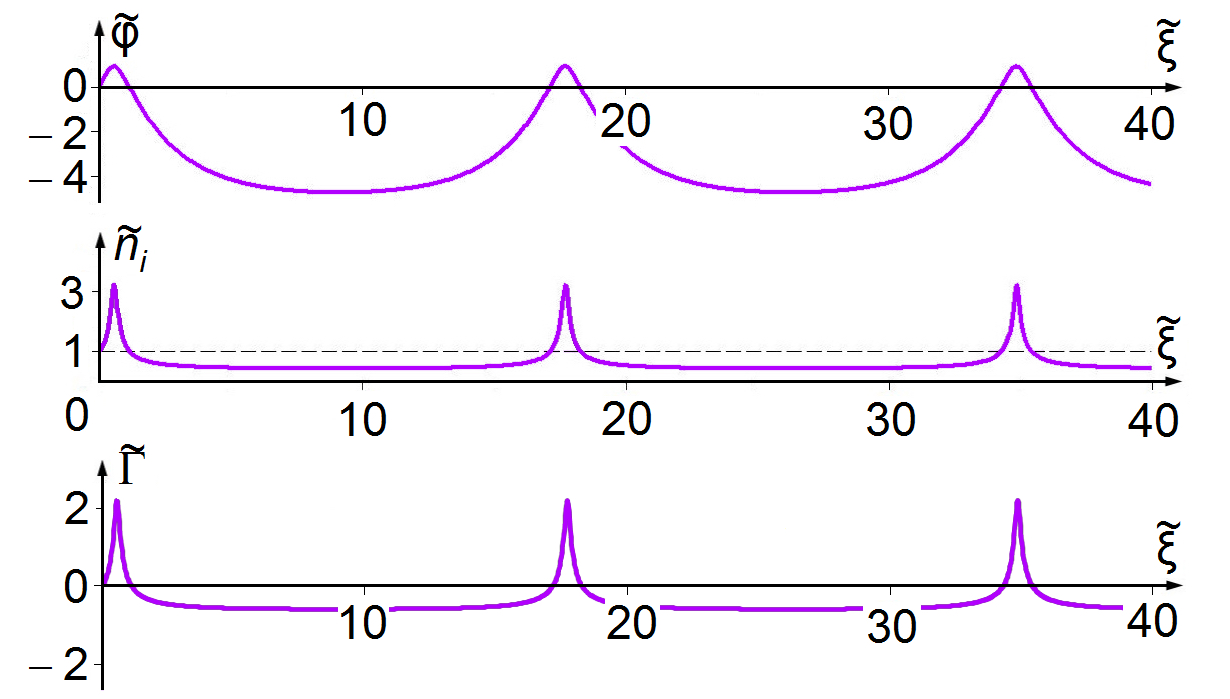}
	\end{center}
	\caption{Oscillation profiles of the \textit{ei}-plasma parameters perturbed by a small amplitude harmonic ion-acoustic wave (left) and by a high-amplitude nonlinear ion-acoustic wave (right). The physical quantities are normalised as $\tilde{\varphi}=-2e\varphi/(m_iV^2)$, $\tilde{n}_i=n_i/n_{0i}$, $\tilde{\Gamma}=\Gamma/(n_{0i}V)$, $\tilde{\xi}=\xi/\lambda_{De}$.
	}
	\label{fig:4+5}
\end{figure*}

{Adopting a multi-fluid approach which allows for taking different populations of charged particles into account,} we calculate and analyse ion fluxes in plane ion-acoustic waves as functions of the initial perturbation amplitude for three plasma models: \textit{ei}-plasma with cold ions (Sec.~\ref{sec:ei}), \textit{eii}-plasma with two types of warm positively charged ions (Sec.~\ref{sec:eii}), and a dusty \textit{eiid}-plasma (Sec.~\ref{sec:eiid}).
{For the bi-ion plasma models (Sec.~\ref{sec:eii} and \ref{sec:eiid}), we demonstrate the possibility for the separation of ion populations of different types by a slow-mode nonlinear ion-acoustic wave.}
{The interest to the plasma of such a complex composition is connected with its omnipresence in the laboratory and natural plasma environments. For example, in the laboratory experiments, the multi-ion plasma can be created by the ignition of gas mixtures by electric discharges (see e.g. Ref.~\onlinecite{2012HEC...46.349D}, and references therein). The subsequent injection of macroscopic particles such as dust and/or aerosols leads to the formation of a dusty \textit{eiid}-plasma (see e.g. Chap.~11 in Ref.~\onlinecite{2011book.....F}, for a more detailed discussion of the dusty plasma formation and physics and technology of discharges). Likewise, electric discharges in natural gas mixtures such as air are also known to generate plasma with multiple ion populations. Moreover, if the discharge occurs in a dusty atmosphere, the dust particles also get electrified and thus form the \textit{eiid}-plasma (see e.g. Ref.~\onlinecite{2016GeoRL..43.4221C}, for the illustration of this process in volcanic eruptions). Another natural multi-component plasma system also often considered to host nonlinear ion-acoustic waves is cometary tails which usually consist of several types of ions, electrons, and dust (see e.g. Ref.~\onlinecite{Bedeir_2021}, for a recent work).}

\section{Ion flux in a nonlinear ion-acoustic wave in a two-component plasma with cold ions (Sagdeev's model)}
\label{sec:ei}

\begin{figure}
	\begin{center}
		\includegraphics[width=0.8\linewidth]{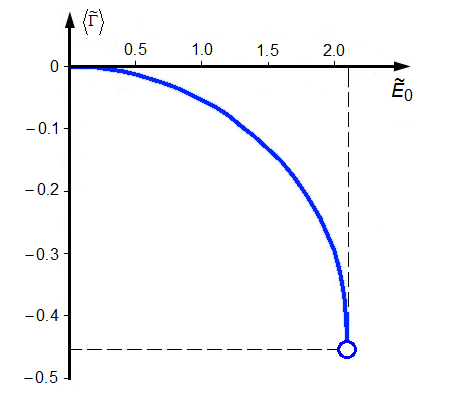}
	\end{center}
	\caption{Dependence of the average wave-induced ion flux $\langle \Gamma\rangle$ (\ref{eq:ei_flux}) in the \textit{ei}-plasma on the electric field $E_0$ perturbing the plasma, normalised as $\tilde{\langle \Gamma\rangle}=\langle \Gamma\rangle/(n_{0i}V)$ and $\tilde{E_0}=-({d\varphi}/{d\xi})|_{\xi=0}2e\lambda_{De}/(m_iV^2)$.
	}
	\label{fig:6}
\end{figure}

\begin{figure}
	\begin{center}
		\includegraphics[width=0.8\linewidth]{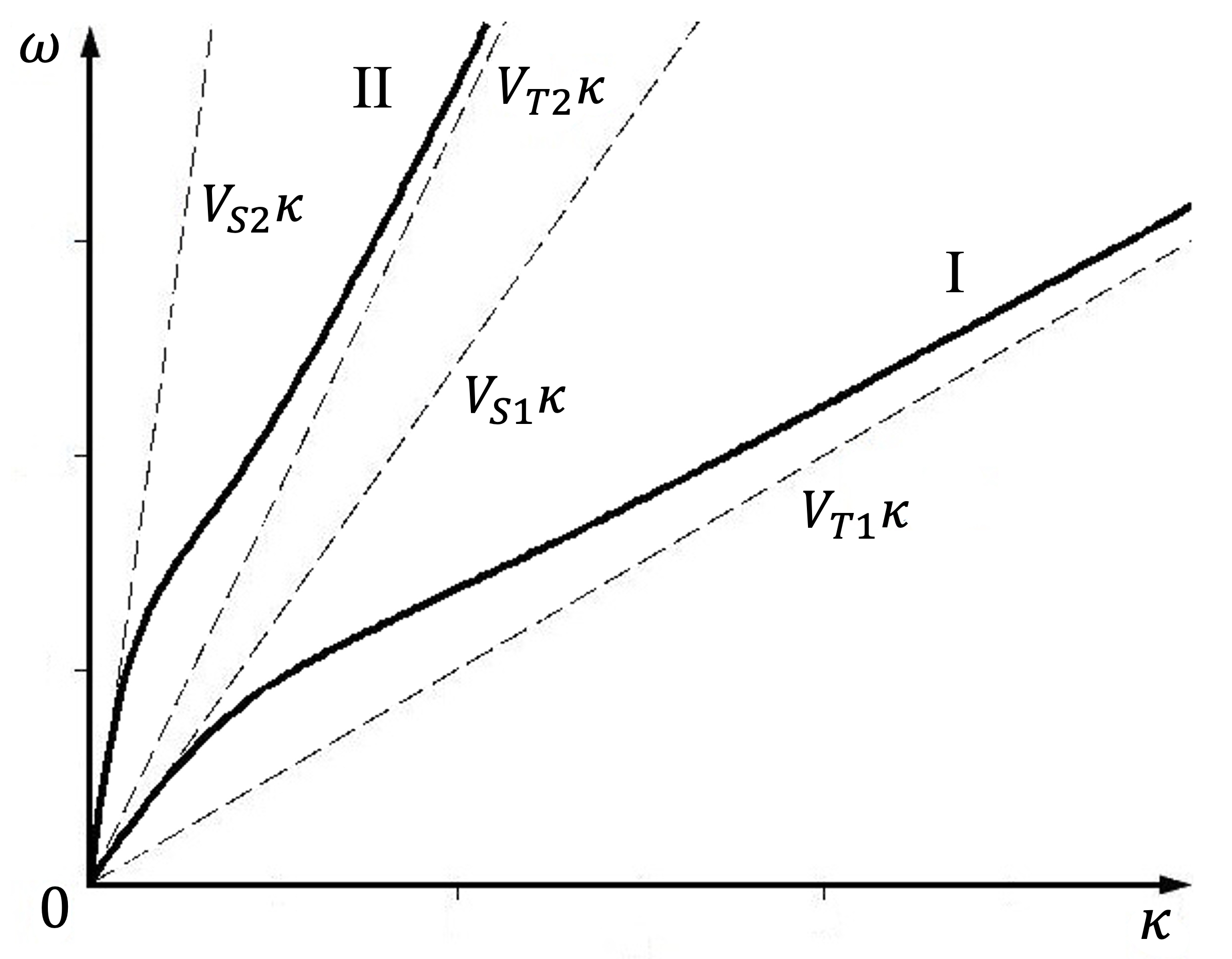}
	\end{center}
	\caption{A schematic sketch of the dispersion relation $\omega(\kappa)$ of ion-acoustic waves in the bi-ion \textit{eii}-plasma, described by Eq.~(\ref{eq:eii_disp}). The short-wavelength and long-wavelength ion-acoustic wave speeds $V_{T1,2}$ and $V_{S1,2}$ are given in Eqs.~(\ref{eq:eii_disp}) and (\ref{eq:eii_vs}), respectively.
	{The Roman numerals \lq\lq I\rq\rq\ and \lq\lq II\rq\rq\ denote the dispersion curves of the linear slow and fast ion-acoustic modes, respectively.}
	}
	\label{fig:7}
\end{figure}

\begin{figure*}
	\begin{center}
		\includegraphics[width=0.48\linewidth]{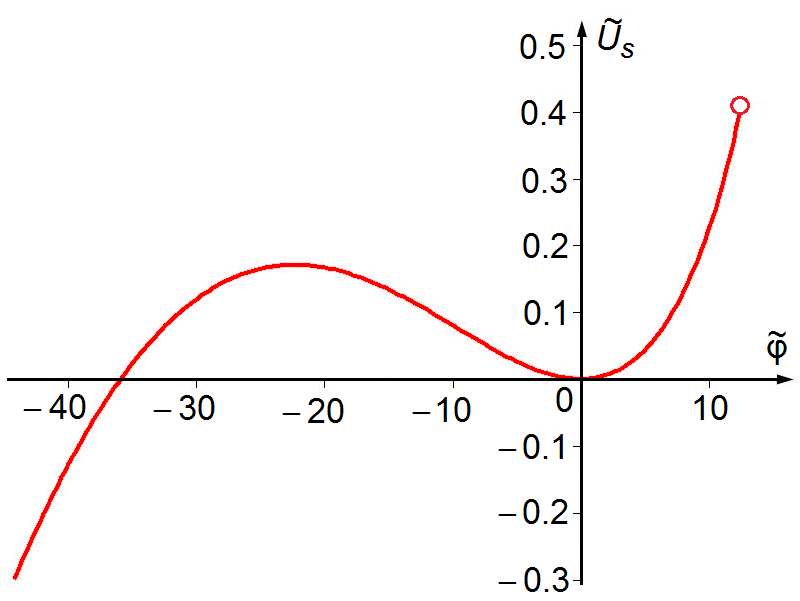}
		\includegraphics[width=0.49\linewidth]{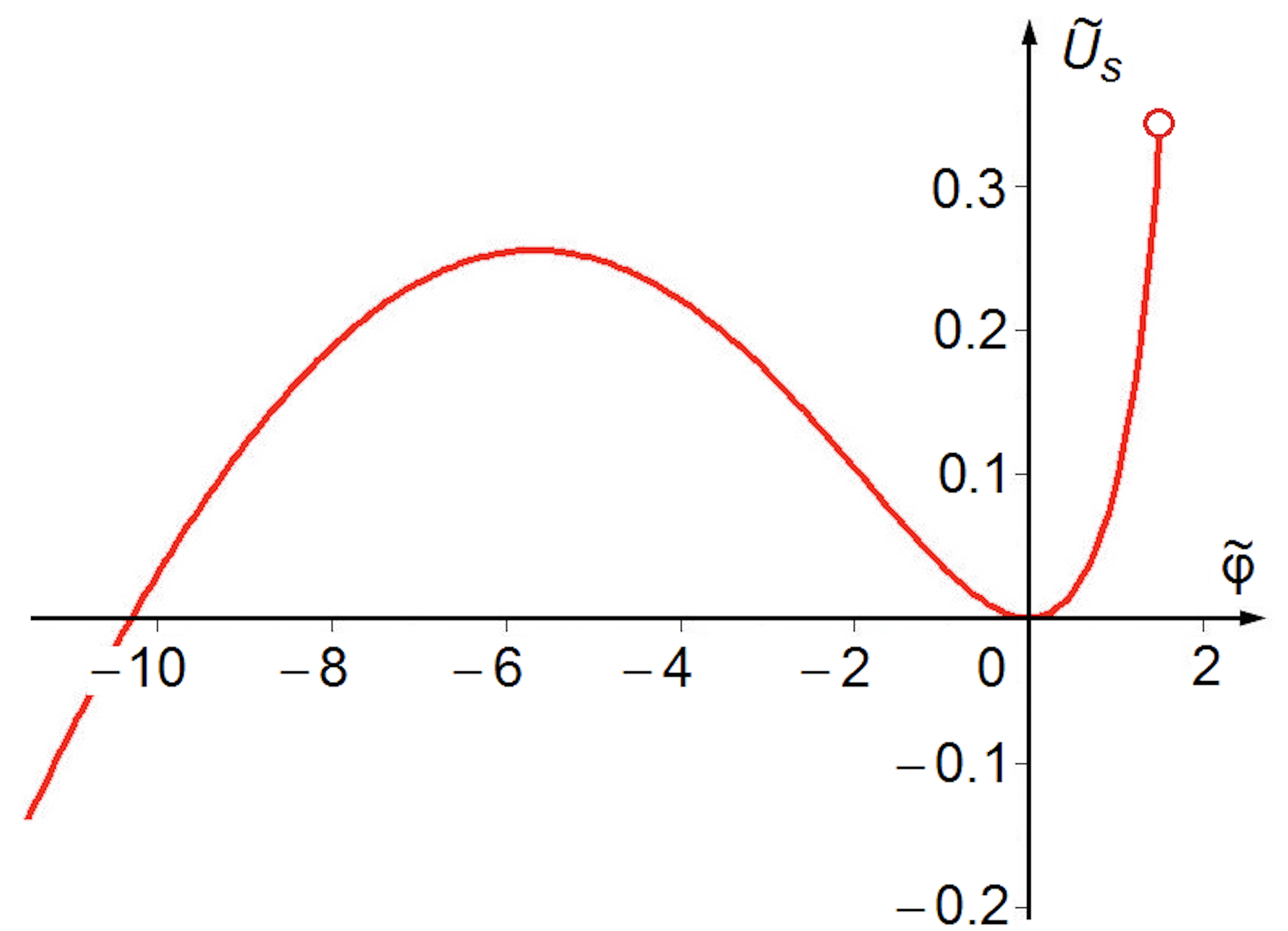}
	\end{center}
	\caption{The Sagdeev pseudopotential $U_S(\varphi)$ obtained for the electric charge density $\rho_{eii}(\varphi)$ (\ref{eq:eii_rho(phi)}) in the bi-ion \textit{eii}-plasma, for the fast (left) and slow (right) ion-acoustic wave mode. The variables are normalised as $\tilde{U}_S=U_S/(2\pi n_{0i1}m_{i1}V^2)$ and $\tilde{\varphi}=-2e\varphi/(m_{i1}V^2)$.
	}
	\label{fig:8+12}
\end{figure*}

\begin{figure*}
	\begin{center}
		\includegraphics[width=0.49\linewidth]{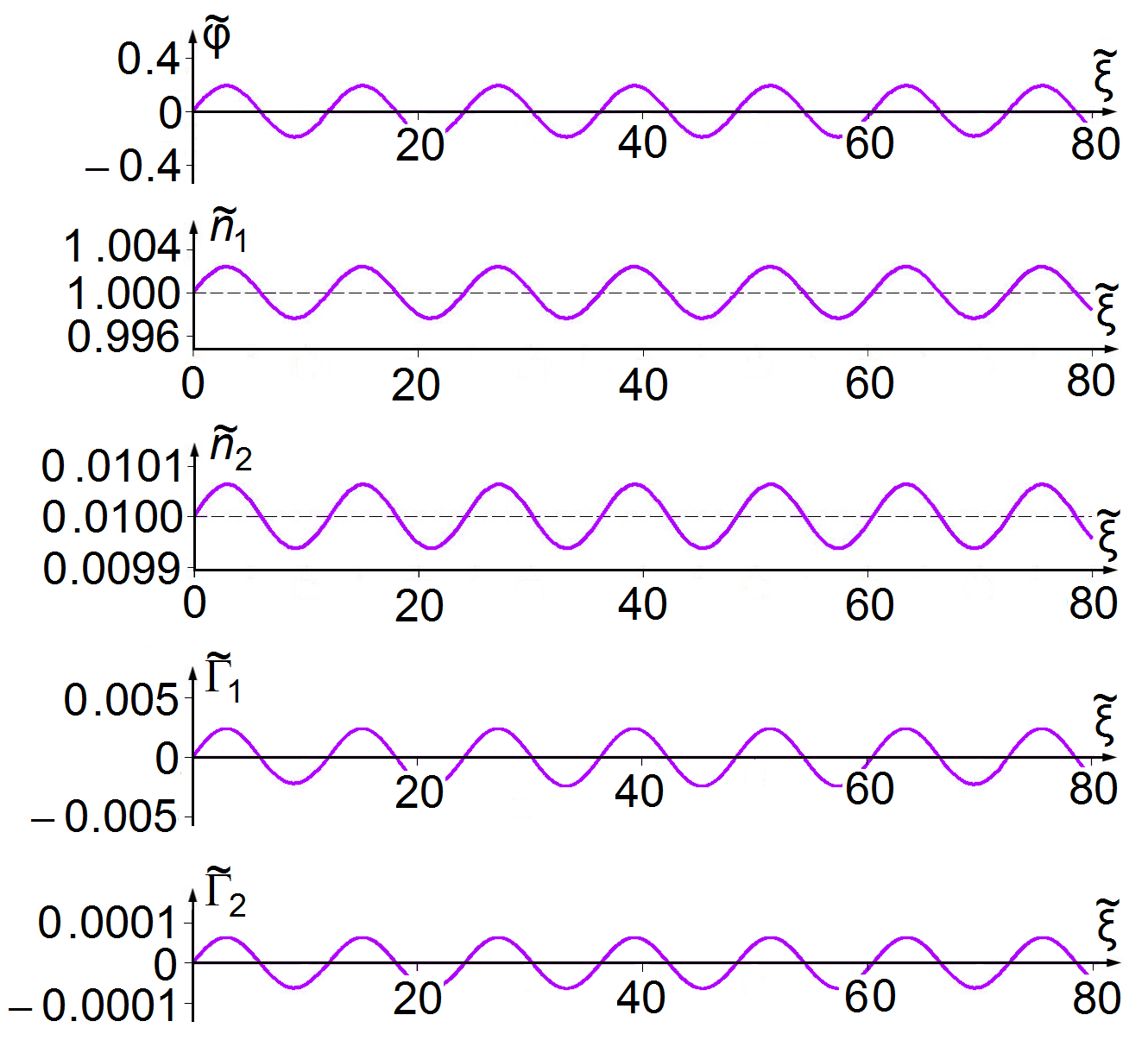}
		\includegraphics[width=0.48\linewidth]{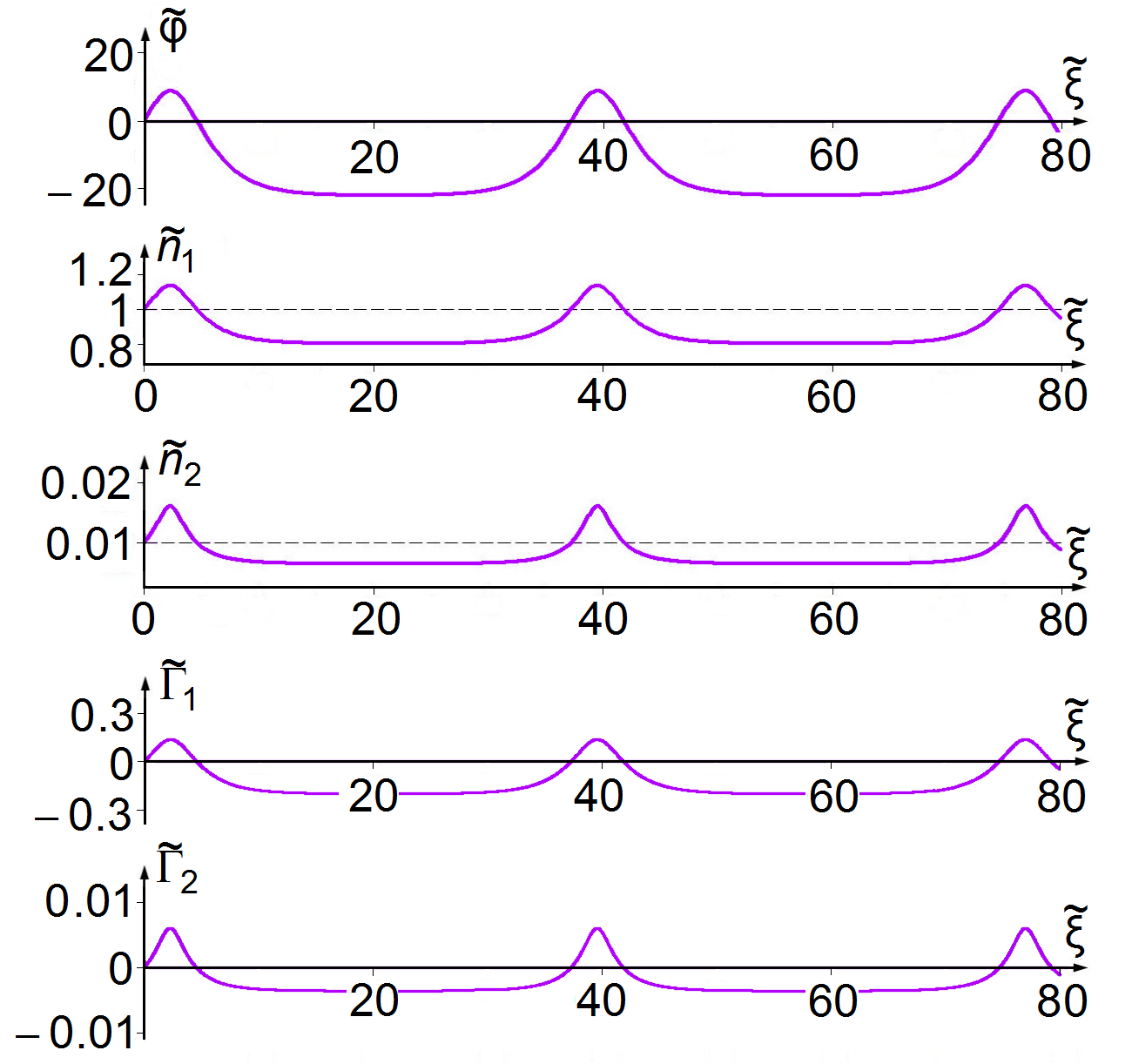}
	\end{center}
	\caption{The same as shown in Fig.~\ref{fig:4+5}, but for the fast ion-acoustic wave in the bi-ion \textit{eii}-plasma. The unperturbed parameters of the ion component \lq\lq 1\rq\rq\ were used for normalisation.
	}
	\label{fig:9+10}
\end{figure*}

An exact solution for the structure of nonlinear ion-acoustic waves in a uniform, collisionless, and unmagnetised \textit{ei}-plasma with cold ions and Boltzmann electrons was derived and analysed by the method of mechanical analogy (also known as the Sagdeev pseudopotential method) in Ref.~\onlinecite{1966RvPP....4...23S}. In this section, we demonstrate the existence of a non-zero average ion-acoustic flux in this model in a nonlinear regime.

The dynamics of ions in such a plasma model is governed by the following set of equations,
\begin{align}\label{eq:ei_cont}
&\frac{\partial n_{i}}{\partial t}+\frac{\partial (n_{i}v_{i})}{\partial x}=0,\\
\label{eq:ei_motion}
&\frac{\partial v_{i}}{\partial t}+v_{i}\frac{\partial v_{i}}{\partial x} = \frac{Ze}{m_{i}}\frac{\partial \varphi}{\partial x},\\
\label{eq:ei_poisson}
&\frac{\partial^2 \varphi}{\partial x^2}=4\pi e(Zn_{i}-n_e),
\end{align}
with the full neutrality condition of the unperturbed plasma $Zen_{0i}-en_{0e}=0$, written for the equilibrium values of the electron and ion concentrations $n_{0e}$ and $n_{0i}$, and the Boltzmann law for the inertialess hot electrons,
\begin{equation}\label{eq:ei_bol_elec}
n_e=n_{0e}\exp\left(-\frac{e\varphi}{k_BT_e}\right).
\end{equation}

Introducing a new variable $\xi$ and transforming to the new frame of reference that moves with the wave phase speed $V$ in the positive direction of the $x$-axis,
\begin{equation}\label{eq:ei_xi}
	\xi = x - Vt,~~~~u_i=v_i-V,
\end{equation}
Eqs.~(\ref{eq:ei_cont})--(\ref{eq:ei_poisson}) can be re-written as
\begin{align}\label{eq:ei_cont2}
&\frac{d(n_{i}u_{i})}{d\xi}=0,\\
\label{eq:ei_motion2}
&u_{i}\frac{d\,u_{i}}{d\,\xi}=\frac{Ze}{m_{i}}\frac{d\varphi}{d\xi},\\
\label{eq:ei_poisson2}
&\frac{d^2 \varphi}{d\,\xi^2}=4\pi e(Zn_{i}-n_e).
\end{align}
Integrating Eqs.~(\ref{eq:ei_cont2})--(\ref{eq:ei_motion2}) with $\lim\limits_{u_i\to-V}n_i=n_{0i}$ and $\lim\limits_{u_i\to-V}\varphi=0$, we obtain explicit nonlinear dependencies between the perturbations of the ion concentration $n_i$, electrostatic potential $\varphi$, and the plasma velocity in the moving frame of reference $u_i$,
\begin{align}
	&u_i = -\frac{n_{0i}}{n_i}V,\label{eq:ei_u(n)}\\
	&n_{i}=n_{0i}\left(1 + \frac{2Ze\varphi}{m_{i}V^2}\right)^{-1/2}.\label{eq:ei_n(phi)}
\end{align}
Using Eqs.~(\ref{eq:ei_bol_elec}) and (\ref{eq:ei_n(phi)}) in Eq.~(\ref{eq:ei_poisson2}) gives the following second-order ordinary differential equation for $\varphi(\xi)$,
\begin{equation}\label{eq:ei_ode}
\frac{d^2 \varphi}{d\,\xi^2}=4\pi\rho_{ei}(\varphi),
\end{equation}
where
\begin{equation}
	\rho_{ei}(\varphi)=Zen_{0i}\left(1 + \frac{2Ze\varphi}{m_{i}V^2}\right)^{-1/2}-en_{0e}\exp\left(-\frac{e\varphi}{k_BT_e}\right).\nonumber
\end{equation}

Following Refs.~\onlinecite{1966RvPP....4...23S, 1979RvMP...51...11S}, Eq.~(\ref{eq:ei_ode}) can be interpreted as the equation of motion of a pseudoparticle in the force field $4\pi\rho_{ei}(\varphi)$, with the variables $\xi$ and $\varphi$ playing the roles of pseudotime and pseudocoordinate, respectively. In this mechanical analogy, the first integral of Eq.~(\ref{eq:ei_ode}) has a form of the conservation of energy law,
\begin{equation}\label{eq:ei_energy}
	\frac{1}{2}\left(\frac{d\varphi}{d\xi}\right)^2+U_S(\varphi)=\mathrm{const},
\end{equation}
with $U_S(\varphi)$ representing the generalised potential energy of the system, also referred to as the Sagdeev pseudopotential,
\begin{align}\label{eq:ei_sagdeev}
&U_S(\varphi)=-4\pi\int_0^{\varphi}\rho_{ei}(\varphi)\,d\varphi= \\
&4\pi m_{i}V^2\Bigg\{n_{0i}\Bigg[1-\bigg(1+\frac{2Ze\varphi}{m_{i}V^2}\bigg)^{1/2}\Bigg]+n_{0e}\frac{k_BT_e}{m_{i}V^2}\left[1-\exp\left(-\frac{e\varphi}{k_BT_e}\right)\right]\Bigg\}.\nonumber
\end{align}
{Without loss of generality, the value of the Sagdeev pseudopotential $U_S(\varphi)$ (\ref{eq:ei_sagdeev}) at $\varphi=0$ is fixed to zero, that implies the lack of the oscillation energy at equilibrium. The perturbation of this equilibrium is introduced through a non-zero value of the parameter $E_0=({d\varphi}/{d\xi})|_{\xi=0}$, which appears as the initial condition for Eq.~(\ref{eq:ei_ode}), has units of the electric field perturbing the plasma, and unambiguously characterises the oscillation energy. The examples of the energy levels corresponding to small-amplitude quasi-linear (low values of $E_0$) and high-amplitude nonlinear (high values of $E_0$) periodic oscillations in the potential field $U_S(\varphi)$ (\ref{eq:ei_sagdeev}) are shown in Fig.~\ref{fig:2+3}.}

The left-hand panel of Fig.~\ref{fig:2+3} shows the Sagdeev pseudopotential $U_S(\varphi)$ (\ref{eq:ei_sagdeev}) for $Z=1$ and $\sqrt{m_iV^2/k_BT_e}=0.6$. It has a standard form with a potential well in which periodic oscillations of $\varphi(\xi)$ can exist. Using Eq.~(\ref{eq:ei_n(phi)}), the Sagdeev pseudopotential (\ref{eq:ei_sagdeev}) with a pseudocoordinate $\varphi$ can be recalculated to another pseudopotential $U_B(n_i)$ with a pseudocoordinate $n_i$ (see Ref.~\onlinecite{2009book.....D}, for details), which is shown in the right-hand panel of Fig.~\ref{fig:2+3}. A similar approach treating the magnetic field and plasma concentration as pseudocoordinates was used by Ref.~\onlinecite{2016PhRvL.117w5102H} for the analysis of nonlinear waves in the terrestrial quasi-parallel foreshock, and by Ref.~\onlinecite{2016PhRvE..93e3205K} for modelling nonlinear oscillations of a current sheet formed between two coalescing magnetic flux ropes.

{The oscillation profiles of the plasma parameters in the small-amplitude linear regime and high-amplitude nonlinear regime are shown by the left-hand and right-hand columns of Fig.~\ref{fig:4+5}, respectively.}
In Fig.~\ref{fig:2+3}, both pseudopotentials $U_S(\varphi)$ and $U_B(n_i)$ can be approximated by a parabolic function near the very bottom of the potential well. Hence, small-amplitude oscillations of $\varphi(\xi)$ described by nonlinear Eq.~(\ref{eq:ei_ode}) and $n_i(\xi)$ obtained from Eq.~(\ref{eq:ei_n(phi)}) are almost indistinguishable from the harmonic linear oscillations (see the left-hand column of Fig.~\ref{fig:4+5}). However, for larger-amplitude perturbations, the potential well in both $U_S(\varphi)$ and $U_B(n_i)$ becomes strongly asymmetric which makes the nonlinear oscillation profiles of $\varphi(\xi)$ and $n_i(\xi)$ highly anharmonic (see the right-hand column of Fig.~\ref{fig:4+5}). In this work, both small-amplitude linear and high-amplitude nonlinear solutions of Eq.~(\ref{eq:ei_ode}) are obtained numerically using the routine \textit{dsolve} in the computing environment \textit{Maple}.

Using Eq.~(\ref{eq:ei_u(n)}) to obtain the oscillation profile of the plasma velocity in the wave frame of reference, $u_i(\xi)$, and in the laboratory frame of reference, $v_i(\xi)=u_i(\xi)+V$, one can calculate the wave-caused variations of the ion mass flux $\Gamma(\xi)$ and average flux $\langle \Gamma\rangle$ over the wave period $\Lambda$ as
\begin{equation}\label{eq:ei_flux}
\Gamma(\xi) = n_i(\xi)v_i(\xi),~~~\langle \Gamma\rangle = \frac{1}{\Lambda}\int_0^{\Lambda}\Gamma(\xi)d\xi.
\end{equation}
Small-amplitude harmonic and high-amplitude strongly anharmonic oscillation profiles of $\Gamma(\xi)$ are shown by Fig.~\ref{fig:4+5}. Figure~\ref{fig:6} shows the dependence of the average ion flux $\langle \Gamma\rangle$ on the parameter $E_0=({d\varphi}/{d\xi})|_{\xi=0}$, {which characterises the initial perturbation energy for Eq.~(\ref{eq:ei_ode}) and considered as a free parameter in this work for discriminating between linear and nonlinear regimes of ion-acoustic waves (see the discussion after Eq.~(\ref{eq:ei_sagdeev})).}

It can be seen from Fig.~\ref{fig:6} that the average flux $\langle \Gamma\rangle$ is non-zero and negative for all values of the parameter $E_0$, except the regime with $E_0\to 0$ for which $\langle \Gamma\rangle \to 0$. The latter means that small-amplitude harmonic waves do not generate the flux, which is consistent, in particular, with the results of Ref.~\onlinecite{1997JTePh..42..713B}, who found no magnetic field in harmonic longitudinal electrostatic waves in plasma. On the other hand, the detected presence of the flux for higher-amplitude perturbations agrees with Ref.~\onlinecite{2000PhPl....7.5252K} reporting on the existence of the magnetic field in nonlinear longitudinal electrostatic waves. No changes in the flux sign similar to those described in Refs.~\onlinecite{2007PhPl...14b2106T, 2012PhPl...19j3702J} were found.

Thus, we demonstrated that nonlinear ion-acoustic waves in the commonly accepted Sagdeev's \textit{ei}-plasma model cause a non-zero average ion flux, counter-directed to the wave propagation. The generation of the flux is attributed to the asymmetry of the generalised potential function $U_B(n_i)$ at high perturbation amplitudes.

\section{Separation of ions by nonlinear ion-acoustic fluxes in a bi-ion plasma}
\label{sec:eii}

\begin{figure}
	\begin{center}
		\includegraphics[width=0.8\linewidth]{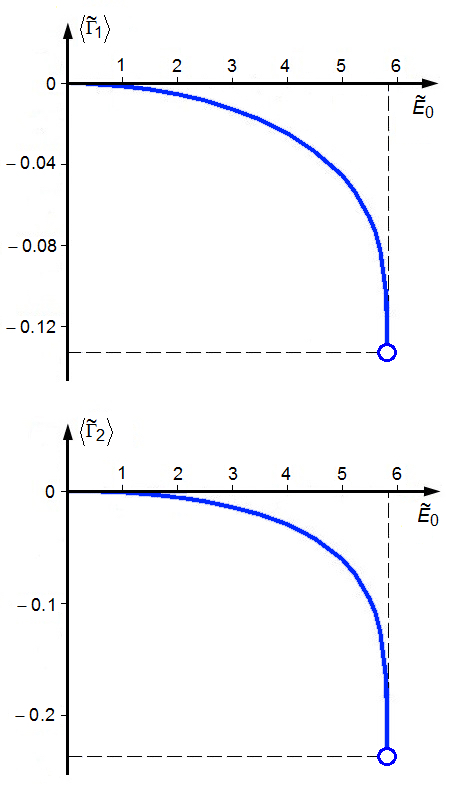}
	\end{center}
	\caption{The same as shown in Fig.~\ref{fig:6}, but for the fast ion-acoustic wave in the bi-ion \textit{eii}-plasma. The unperturbed parameters of the ion component \lq\lq 1\rq\rq\ were used for normalisation.
	}
	\label{fig:11}
\end{figure}

In a bi-ion uniform, collisionless, and unmagnetised plasma with two types of isothermal and positively charged ions and hot electrons, {typical for, for example, auroral regions\cite{1991P&SS...39.1343R}}, the dynamics of ion-acoustic waves is governed by
\begin{align}\label{eq:eii_cont}
	&\frac{\partial n_{i1,2}}{\partial t}+\frac{\partial (n_{i1,2}v_{i1,2})}{\partial x}=0,\\
	\label{eq:eii_motion}
	&\frac{\partial v_{i1,2}}{\partial t}+v_{i1,2}\frac{\partial v_{i1,2}}{\partial x} = \frac{Z_{1,2}e}{m_{i1,2}}\frac{\partial \varphi}{\partial x} - \frac{1}{m_{i1,2}n_{i1,2}}\frac{\partial P_{i1,2}}{\partial x},\\
	\label{eq:eii_poisson}
	&\frac{\partial^2 \varphi}{\partial x^2}=4\pi e(Z_1n_{i1}+Z_2n_{i2}-n_e),\\
	\label{eq:eii_state}
	&P_{i1,2}=n_{i1,2}k_BT_{i1,2}.
\end{align}
The condition of a full initial neutrality for this plasma model takes the form $Z_1en_{0i1}+Z_2en_{0i2}-en_{0e}=0$, and the electron concentration $n_e$ is considered to obey Boltzmann law (\ref{eq:ei_bol_elec}). The subscripts \lq\lq 1\rq\rq\ and \lq\lq 2\rq\rq\ stand for the ions of different types.

\subsection{Linear theory}\label{sec:eii_lin}

For the analysis of the linear regime of the isothermal ion-acoustic waves described by Eqs.~(\ref{eq:eii_cont})--(\ref{eq:eii_state}), we write all the oscillating variables in the form of harmonic functions $f=f_0+f_1\exp[i(\kappa x-\omega t)]$ with a small amplitude $f_1$, the wavenumber $\kappa$, and cyclic frequency $\omega$. After linearisation, this gives the dispersion relation,
\begin{equation}\label{eq:eii_disp}
	1=-\frac{1}{\lambda_{De}^2\kappa^2}+\frac{\omega_1^2}{\omega^2-V_{T1}^2\kappa^2}+\frac{\omega_2^2}{\omega^2-V_{T2}^2\kappa^2},
\end{equation}
where $V_{T1,2}^2=k_BT_{i1,2}/m_{i1,2}$ are the ion thermal speeds squared (i.e. isothermal sound speeds in the short-wavelength limit when the effect of dispersion is negligible), $\omega_{1,2}^2=4\pi(Z_{1,2}e)^2n_{0i1,2}/m_{i1,2}$ are the ion plasma frequencies squared, and $\lambda_{De}^2=k_BT_e/(4\pi e^2n_{0e})$ is the electron Debye length squared {(see e.g. Sec.~3.2 in Ref.~\onlinecite{1985book.....D}, where a similar dispersion relation was derived in more detail)}.

\begin{figure*}
	\begin{center}
		\includegraphics[width=0.49\linewidth]{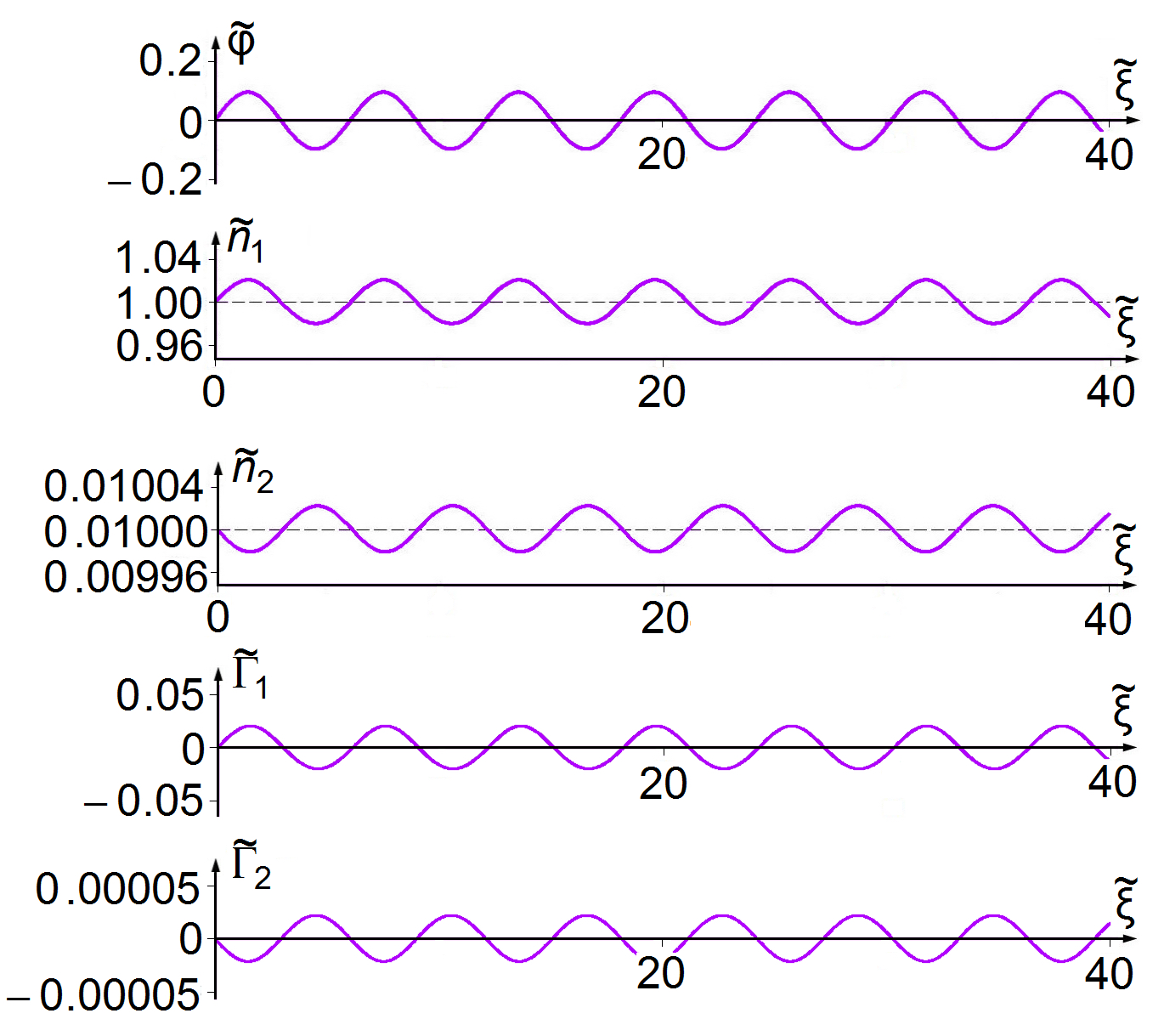}
		\includegraphics[width=0.47\linewidth]{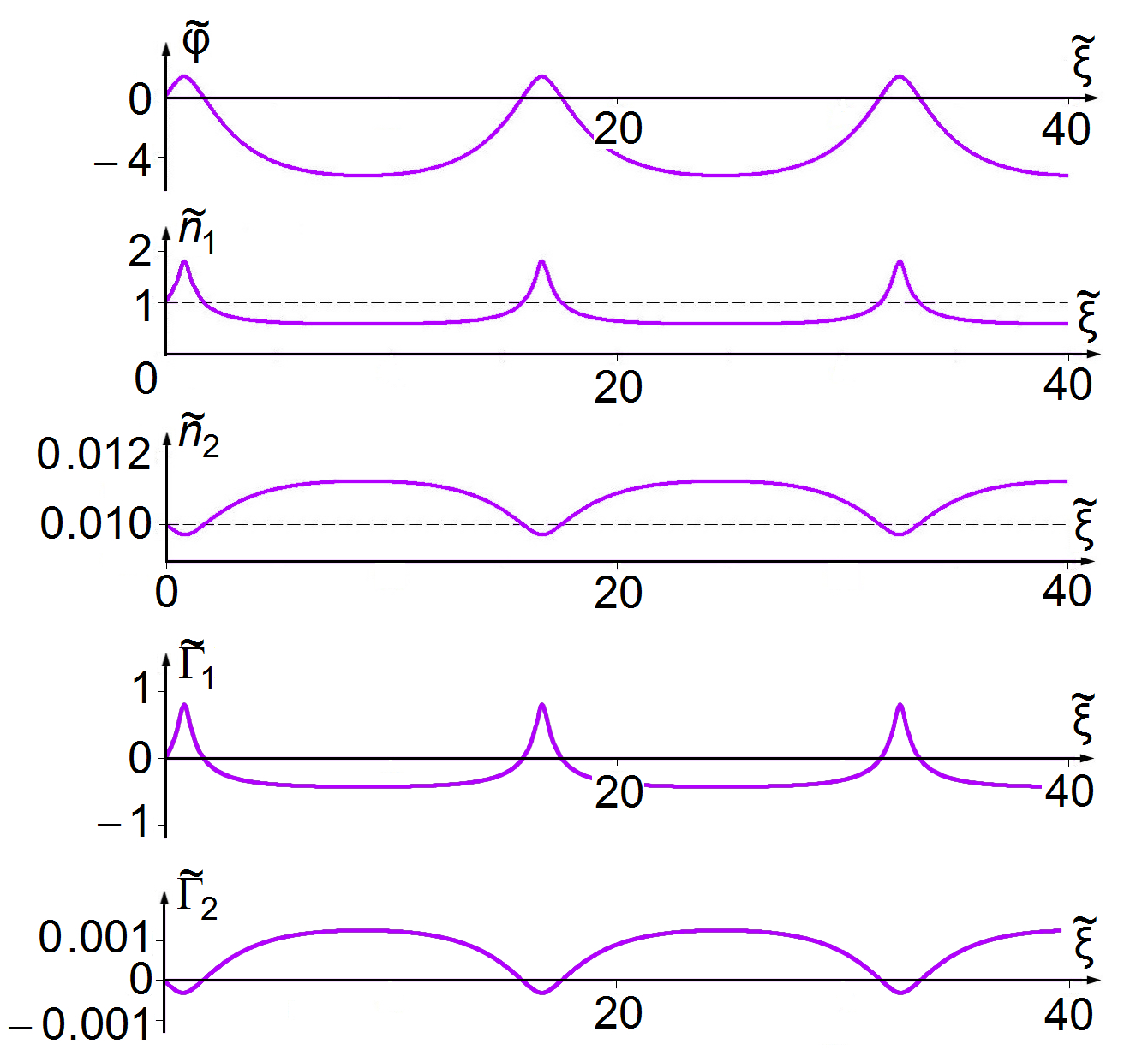}
	\end{center}
	\caption{The same as shown in Fig.~\ref{fig:4+5}, but for the slow ion-acoustic wave in the bi-ion \textit{eii}-plasma. The unperturbed parameters of the ion component \lq\lq 1\rq\rq\ were used for normalisation.
	}
	\label{fig:13+14}
\end{figure*}

The dependence $\omega(\kappa)$ obtained from Eq.~(\ref{eq:eii_disp}) is illustrated in Fig.~\ref{fig:7}. It contains two dispersion curves which correspond to a slow ion-acoustic wave ({the dispersion curve I}) in which ions of different types oscillate in anti-phase, and a fast ion-acoustic wave ({the dispersion curve II}) in which oscillations of ions of different types are in-phase (see also Ref.~\onlinecite{2009PlPhR..35..991D}). As such, long-wavelength slow and fast ion-acoustic waves propagate at different sound speeds which can be determined from the condition $V_S=\lim\limits_{\kappa\to0}{d\omega}/{d\kappa}$ as
\begin{align}
	&V_{S1,2}=\left\{\frac{1}{2}\left[\lambda_{De}^2\left(\omega_1^2+\omega_2^2\right)+\left(V_{T1}^2+V_{T2}^2\right)\right]\pm\frac{1}{2}\left[\left(\omega_1^2+\omega_2^2\right)^2\lambda_{De}^4\right.\right.\nonumber\\
	&\left.\left.+2\left(\omega_1^2-\omega_2^2\right)\left(V_{T1}^2-V_{T2}^2\right)\lambda_{De}^2+\left(V_{T1}^2-V_{T2}^2\right)^2\right]^{1/2}\right\}^{1/2},\label{eq:eii_vs}
\end{align}
where plus/minus corresponds to fast/slow ion-acoustic wave, respectively. In the further analysis, we consider the generation of nonlinear ion fluxes in both these wave modes.
 
\subsection{Nonlinear theory and separation of ions}

Applying transformation to the wave frame of reference (\ref{eq:ei_xi}) to Eqs.~(\ref{eq:eii_cont})--(\ref{eq:eii_state}) and integrating the continuity and motion equations in a similar manner described in Sec.~\ref{sec:ei}, we obtain the explicit dependences between the wave-caused perturbations of the concentrations of the ion components $n_{i1,2}$ and the electrostatic potential $\varphi$,
\begin{align}\label{eq:eii_n12}
	&\frac{n_{i1,2}(\varphi)}{n_{0i1,2}}=\\
	&\left\{-\frac{k_BT_{i1,2}}{m_{i1,2}V^2}W_{0,-1}\left[-\frac{m_{i1,2}V^2}{k_BT_{i1,2}}\exp\left(-\frac{m_{i1,2}V^2}{k_BT_{i1,2}}-\frac{2Z_{1,2}e\varphi}{k_BT_{i1,2}}\right)\right]\right\}^{-1/2}.\nonumber
\end{align}
Solution (\ref{eq:eii_n12}) is written through the Lambert $W$-function of $\varphi$ which is known to have two real branches denoted by the subscripts \lq\lq 0\rq\rq\ and \lq\lq -1\rq\rq (see Ref.~\onlinecite{2005JPlPh..71..715D}). Demanding a full initial neutrality of the plasma and following the recipe developed by Refs.~\onlinecite{2006PhPl...13h2111D, 2012ITPS...40.1429D, 2020JETP..131..844D}, we take the lower real branch $W_{-1}(..)$ for $m_{i1,2}V^2>k_BT_{i1,2}$, and the main real branch $W_{0}(..)$ for $m_{i1,2}V^2<k_BT_{i1,2}$.

Substituting Eqs.~(\ref{eq:ei_bol_elec}) and (\ref{eq:eii_n12}) in Eq.~(\ref{eq:eii_poisson}) written in the moving frame of reference (\ref{eq:ei_xi}), we obtain second-order ordinary differential equation (\ref{eq:ei_ode}) with the electric charge density $\rho_{eii}(\varphi)$ given by
\begin{align}\label{eq:eii_rho(phi)}
&\rho_{eii}(\varphi)=-en_{0e}\exp\left(-\frac{e\varphi}{k_BT_e}\right)\\
&+Z_1en_{0i1}\left\{-\frac{k_BT_{i1}}{m_{i1}V^2}W_{0,-1}\left[-\frac{m_{i1}V^2}{k_BT_{i1}}\exp\left(-\frac{m_{i1}V^2}{k_BT_{i1}}-\frac{2Z_{1}e\varphi}{k_BT_{i1}}\right)\right]\right\}^{-1/2}\nonumber\\
&+Z_2en_{0i2}\left\{-\frac{k_BT_{i2}}{m_{i2}V^2}W_{0,-1}\left[-\frac{m_{i2}V^2}{k_BT_{i2}}\exp\left(-\frac{m_{i2}V^2}{k_BT_{i2}}-\frac{2Z_{2}e\varphi}{k_BT_{i2}}\right)\right]\right\}^{-1/2}\nonumber.
\end{align}

Similarly to Sec.~\ref{sec:ei}, Eq.~(\ref{eq:ei_ode}) with (\ref{eq:eii_rho(phi)}) can be interpreted as a generalised equation of motion of a pseudoparticle in the potential field $U_S(\varphi)=-4\pi\int_0^{\varphi}\rho_{eii}(\varphi)\,d\varphi$, whose explicit form can be obtained analytically by integration of (\ref{eq:eii_rho(phi)}) {as shown} in Appendix \ref{app:eii_sagdeev}. The left-hand panel of Fig.~\ref{fig:8+12} shows the pseudopotential $U_S(\varphi)$ for (\ref{eq:eii_rho(phi)}) and $m_{i2}/m_{i1}=0.4$, $n_{0i2}/n_{0i1}=0.01$, $\sqrt{m_{i1}V^2/k_BT_{i1}}=9$, $Z_1=Z_2=1$, $T_{i1}/T_e$=0.01, and $T_{i2}/T_e=0.02$. For this set of \textit{eii}-plasma parameters, the wave phase speed $V$ corresponds to the fast ion-acoustic wave (appears in region II in Fig.~\ref{fig:7}). We also note that for the chosen set of parameters, one need to use the lower real branch $W_{-1}(..)$ of the $W$-Lambert function in Eq.~(\ref{eq:eii_n12}) for both $n_{i1}$ and $n_{i2}$ to satisfy the condition of a full initial neutrality of the plasma.

Figure~\ref{fig:9+10} shows the oscillation profiles of the electrostatic potential $\varphi(\xi)$, ion concentrations $n_{i1,2}(\xi)$, and also fluxes of both types of ions $\Gamma_1(\xi)$ and $\Gamma_2(\xi)$ calculated by Eq.~(\ref{eq:ei_flux}), for a quasi-harmonic fast ion-acoustic wave with the amplitude near the very bottom of the potential well $U_S(\varphi)$ {(left-hand column of Fig.~\ref{fig:9+10})}, and for strongly nonlinear fast ion-acoustic wave with the amplitude just below the maximum height of the potential barrier {(right-hand column of Fig.~\ref{fig:9+10})}. It is clear from Fig.~\ref{fig:9+10} that both ion populations experience in-phase oscillations, whereas the nonlinear oscillation profiles have a distinct asymmetry in comparison with the linear regime. Using Eq.~(\ref{eq:ei_flux}), we have also calculated mean ion fluxes over the wave period $\langle\Gamma_1\rangle$ and $\langle\Gamma_2\rangle$ for both ion components of the plasma, whose dependence on the perturbation amplitude $E_0=({d\varphi}/{d\xi})|_{\xi=0}$ is shown in Fig.~\ref{fig:11}. Thus, both average ion fluxes $\langle\Gamma_1\rangle$ and $\langle\Gamma_2\rangle$ in the nonlinear fast ion-acoustic wave are found to be negative (i.e. directed against the wave propagation) for all perturbation amplitudes.

\begin{figure}
	\begin{center}
		\includegraphics[width=0.8\linewidth]{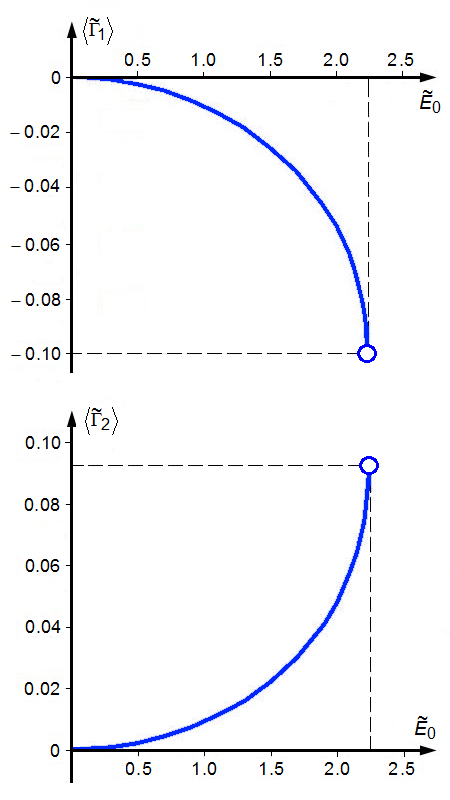}
	\end{center}
	\caption{The same as shown in Fig.~\ref{fig:6}, but for the slow ion-acoustic wave in the bi-ion \textit{eii}-plasma. The unperturbed parameters of the ion component \lq\lq 1\rq\rq\ were used for normalisation.
	}
	\label{fig:15}
\end{figure}

The right-hand panel of Fig.~\ref{fig:8+12} shows the pseudopotential $U_S(\varphi)$ for the slow ion-acoustic wave in the bi-ion plasma, obtained for $m_{i2}/m_{i1}=1$, $n_{0i2}/n_{0i1}=0.01$, $\sqrt{m_{i1}V^2/k_BT_{i1}}=2.4$, $Z_1=Z_2=1$, $T_{i1}/T_e$=0.1, and $T_{i2}/T_e=5$. In this regime, the initial neutrality condition is satisfied by the use of the lower real branch $W_{-1}(..)$ of the $W$-Lambert function for $n_{i1}$ and its main real branch $W_{0}(..)$ for $n_{i2}$ in Eq.~(\ref{eq:eii_n12}). The oscillation profiles of the physical variables, including fluxes $\Gamma_1(\xi)$ and $\Gamma_2(\xi)$ of both ion populations, are shown in Fig.~\ref{fig:13+14}, for a small-amplitude linear and high-amplitude nonlinear regimes {(left-hand and right-hand columns, respectively)}. In both cases, the ions of different types oscillate in anti-phase, which indicates a slow nature of this ion-acoustic wave. The mean fluxes $\langle\Gamma_1\rangle$ and $\langle\Gamma_2\rangle$ for both types of ions in the considered slow ion-acoustic wave are shown in Fig.~\ref{fig:15} against the perturbation amplitude. In contrast to the co-directed fluxes in the fast ion-acoustic wave, mean ion fluxes in the slow wave are directed oppositely to each other, even despite the fact that both ion components have the same sign of the electric charge. In other words, the nonlinear slow ion-acoustic wave causes ions of different types to drift in the opposite directions, which can be used for their effective separation.


\section{Ion fluxes and separation in super-nonlinear ion-acoustic waves in a multi-component plasma with dust grains and warm ions}
\label{sec:eiid}

The revealed ion separation process can be substantially enhanced by the use of a super-nonlinear regime of a slow ion-acoustic wave.
The concept of super-nonlinear ion-acoustic waves was proposed by Refs.~\onlinecite{2012JTePh..57..585D, 2012PlPhR..38..833D}, see also Ref.~\onlinecite{2019Ap&SS.364..180S} for a more recent work. It is based on the existence of two or more local minima separated by local maxima in the Sagdeev pseudopotential, within which the periodic stationary waves with the oscillation amplitude above the separatrix layer are referred to as super-nonlinear (see also Ref.~\onlinecite{2018RvMPP...2....2D} for a comprehensive review and numerous references therein). In practice, this regime can be achieved by adding a heavy dust fraction to the plasma. Thus, in this section we consider a multi-species dusty plasma with electrons, two types of positevely charged ions, and negatively charged static dust grains. As previously, the plasma is assumed to be uniform, collisionless, and unmagnetised, while the electric charge of dust particles is constant.
{The models of the wave-hosting dusty plasmas with a constant dust charge were considered by, for example, Refs.~\onlinecite{1992PhyS...45..508S, 2003NJPh....5...17S}. Likewise, Ref.~\onlinecite{2008JTePh..53.1129D} found the structure of ion-acoustic waves to be weakly affected by the variable charge of the dust particles.}

For such an \textit{eiid}-plasma, the full initial neutrality condition takes the form $Z_1en_{0i1}+Z_2en_{0i2}-en_{0e}-q_dn_d=0$, where the dusty fraction can be described by a dimensionless parameter $\alpha=q_dn_d/en_{0i1}$. For the description of the ion-acoustic wave dynamics in such a plasma model we use the continuity, motion, and state equations (\ref{eq:eii_cont}), (\ref{eq:eii_motion}), and (\ref{eq:eii_state}), while Poisson's equation accounting for the electric charge density of dust becomes
\begin{equation}
\label{eq:eiid_poisson}
\frac{\partial^2 \varphi}{\partial x^2}=4\pi e(Z_1n_{i1}+Z_2n_{i2}-n_e - \alpha n_{0i1}).
\end{equation}
{The model (\ref{eq:eii_cont}), (\ref{eq:eii_motion}), (\ref{eq:eii_state}), and (\ref{eq:eiid_poisson}) does not account for the effect of gravity, which may become important for the dust grains larger than $\sim$10\,$\mu$m and is less pronounced for the dust grains smaller than $\sim$1\,$\mu$m (see e.g. Ref.~\onlinecite{1996P&SS...44..239B}, for more details on the dispensing of dust particles into the plasma and experiments on ion-acoustic waves in the dusty plasma device).}

The linear theory of ion-acoustic waves in the \textit{eiid}-plasma with static dust grains of a constant electric charge is qualitatively similar to that for \textit{eii}-plasma, described in Sec.~\ref{sec:eii_lin}. Namely, there will also be two dispersion curves corresponding to the fast and slow ion-acoustic wave modes, whose characteristic speeds can be found by Eq.~(\ref{eq:eii_vs}). For the nonlinear analysis, the application of the methodology described in Secs.~\ref{sec:ei} and \ref{sec:eii} gives the electric charge density
\begin{equation}
	\label{eq:eiid_rho(phi)}
	\rho_{eiid}(\varphi) = \rho_{eii}(\varphi) - \alpha n_{0i1},
\end{equation}
where $\rho_{eii}(\varphi)$ is given by Eq.~(\ref{eq:eii_rho(phi)}).

\begin{figure}
	\begin{center}
		\includegraphics[width=\linewidth]{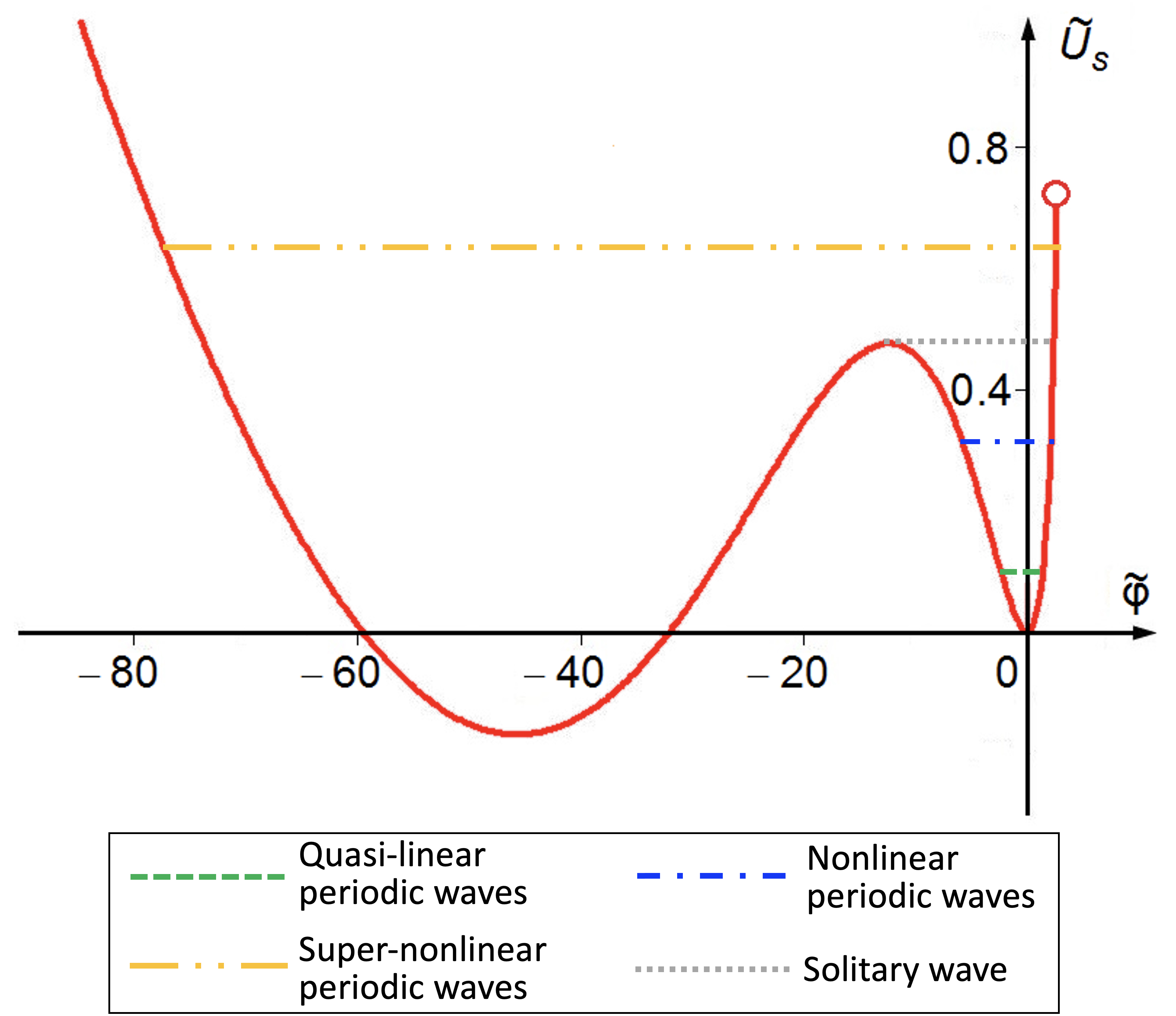}
	\end{center}
	\caption{The Sagdeev pseudopotential $U_S(\varphi)$ obtained for the electric charge density $\rho_{eiid}(\varphi)$ (\ref{eq:eiid_rho(phi)}) in the dusty \textit{eiid}-plasma, for the slow ion-acoustic wave. The variables are normalised as $\tilde{U}_S=U_S/(2\pi n_{0i1}m_{i1}V^2)$ and $\tilde{\varphi}=-2e\varphi/(m_{i1}V^2)$.
	{The examples of the energy levels corresponding to small-amplitude quasi-linear, high-amplitude nonlinear, and large-amplitude super-nonlinear periodic waves around the initial equilibrium $\varphi=0$ are shown in green, blue, and yellow, respectively. The energy level of a solitary wave (not considered in this work) is shown in grey.}
	}
	\label{fig:16+21}
\end{figure}

\begin{figure*}
	\begin{center}
		\includegraphics[width=0.33\linewidth]{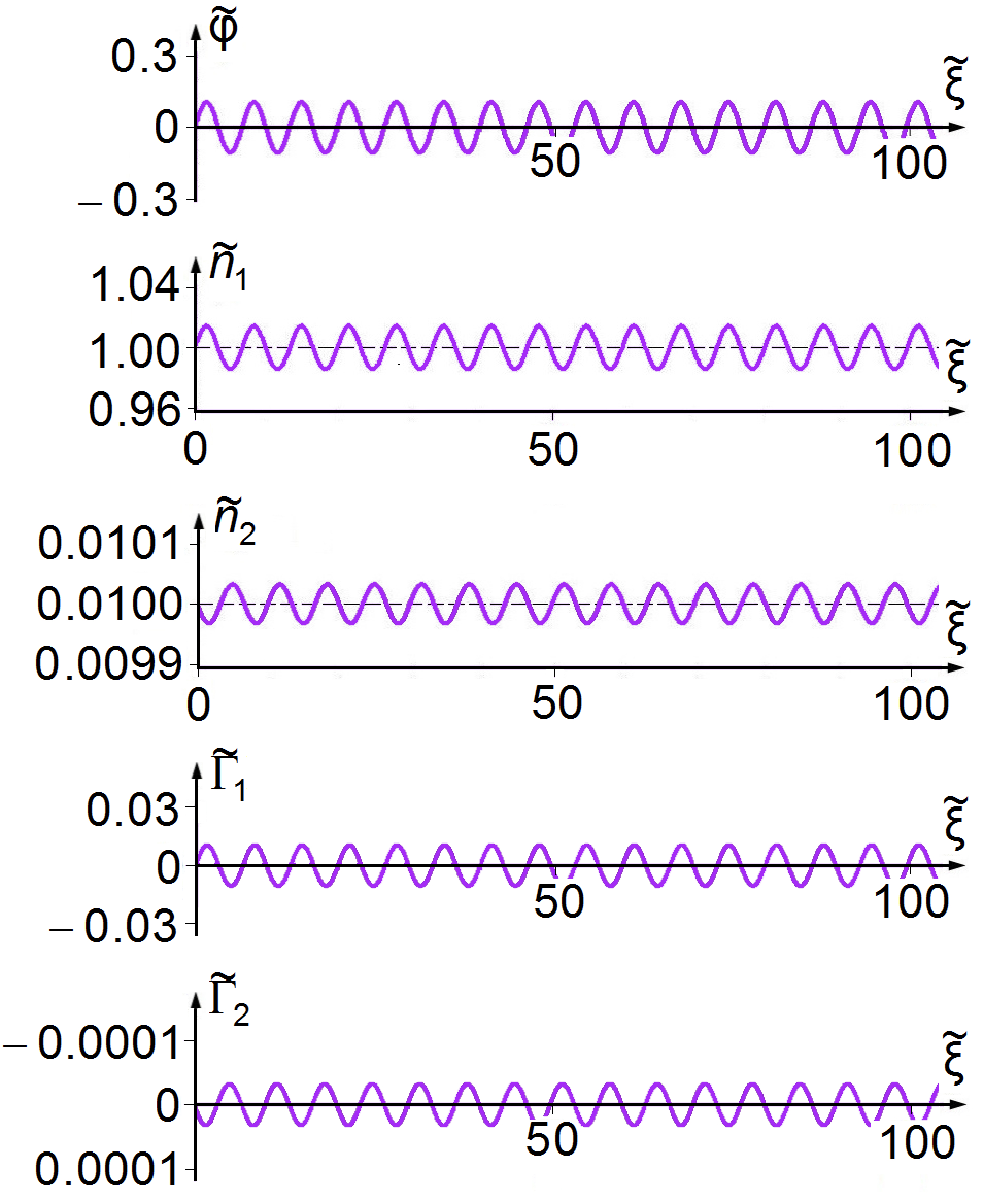}
		\includegraphics[width=0.335\linewidth]{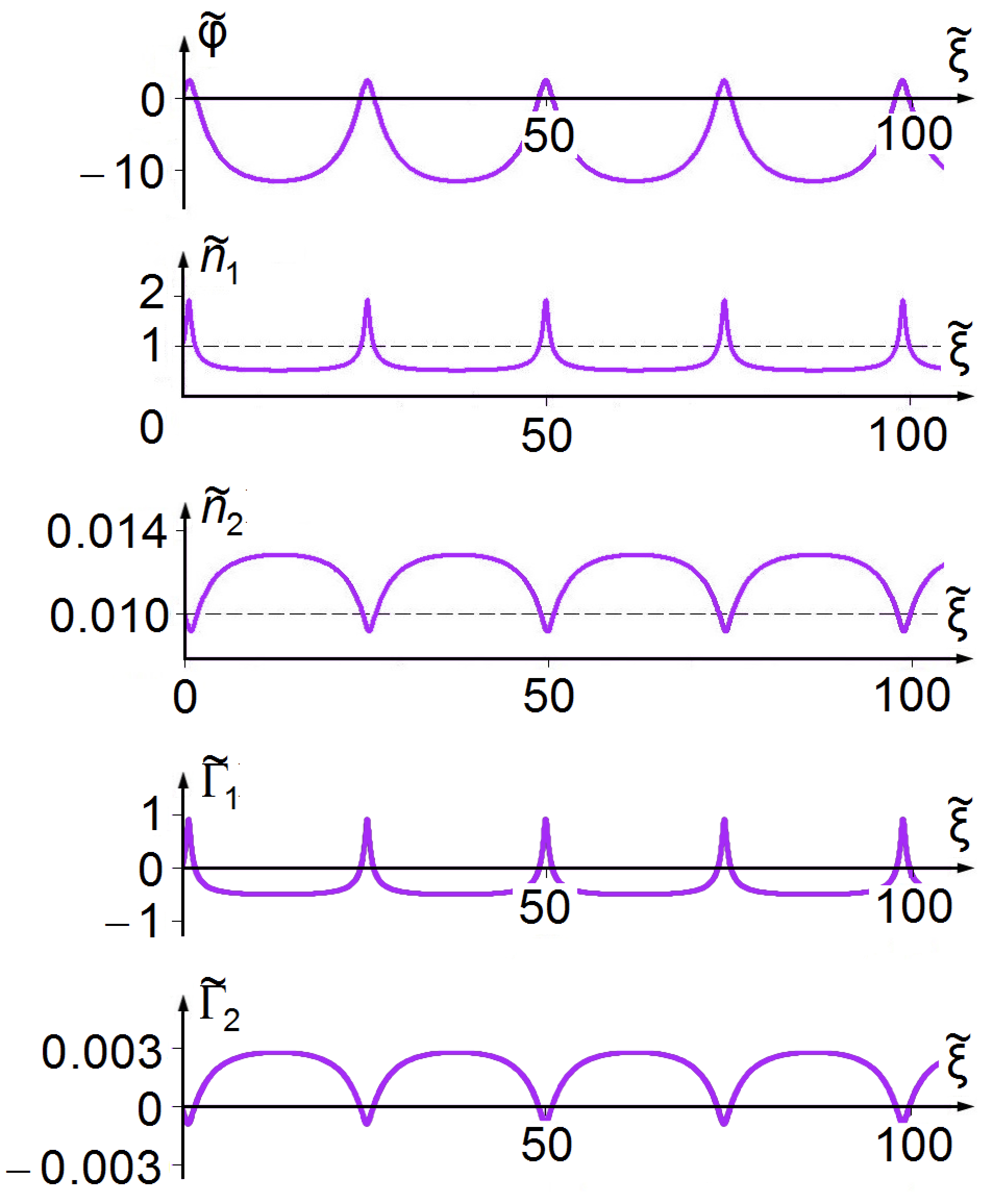}
		\includegraphics[width=0.315\linewidth]{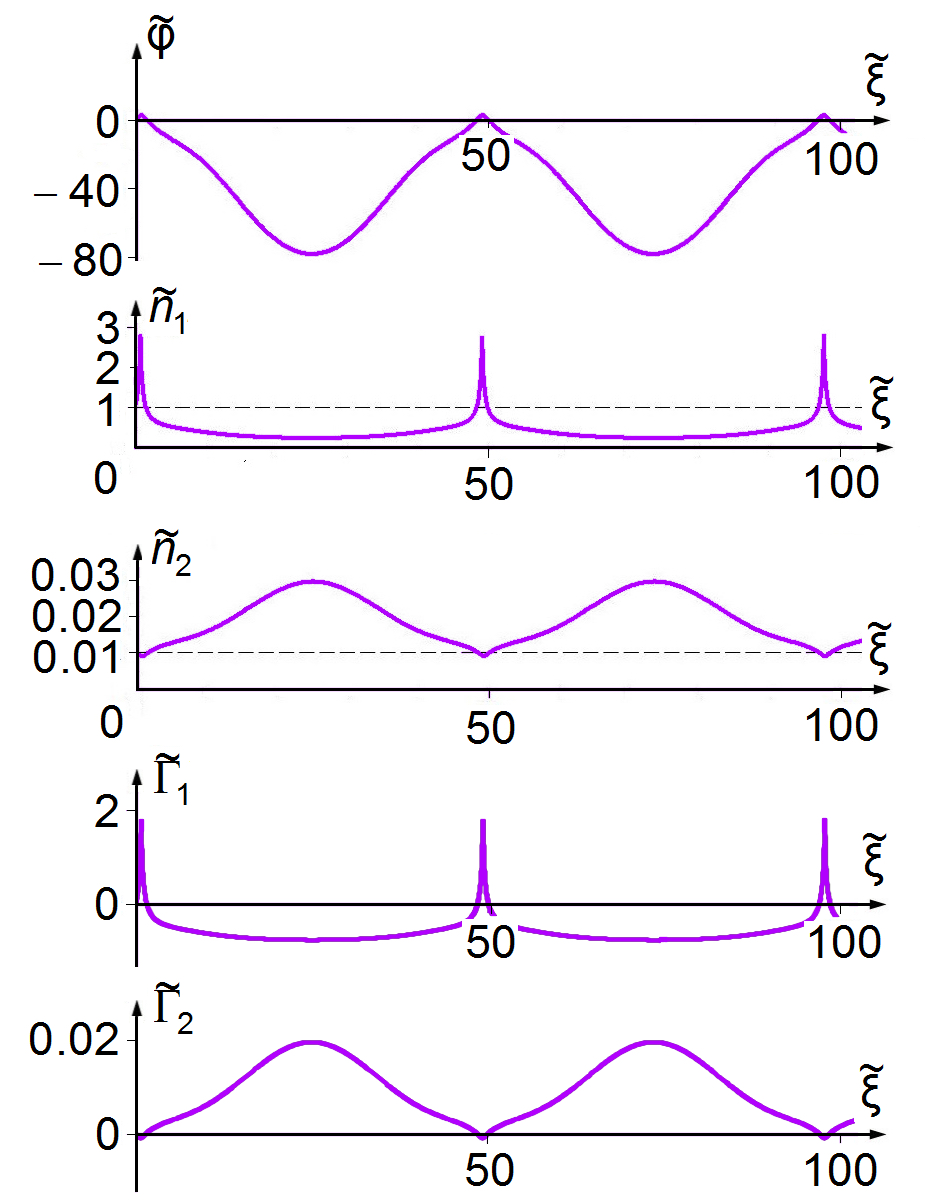}
	\end{center}
	\caption{Oscillation profiles of the \textit{eiid}-plasma parameters perturbed by a slow ion-acoustic wave in the small-amplitude harmonic regime (left), high-amplitude nonlinear regime (middle), and large-amplitude super-nonlinear regime (right). The physical quantities are normalised as $\tilde{\varphi}=-2e\varphi/(m_{i1}V^2)$, $\tilde{n}_{i1,2}=n_{i1,2}/n_{0i1}$, $\tilde{\Gamma}_{1,2}=\Gamma/(n_{0i1}V)$,  $\tilde{\xi}=\xi/\lambda_{De}$.
	}
	\label{fig:22+23+24}
\end{figure*}

The subsequent integration of $\rho_{eiid}(\varphi)$ (\ref{eq:eiid_rho(phi)}) by $\varphi$ gives the Sagdeev pseudopotential $U_S(\varphi)=-4\pi\int_0^{\varphi}\rho_{eiid}(\varphi)\,d\varphi$ for nonlinear ion-acoustic waves in the considered \textit{eiid}-plasma, which is {shown in Appendix \ref{app:eiid_sagdeev} and} illustrated in Fig.~\ref{fig:16+21}. Here, $U_S(\varphi)$ is shown for $m_{i2}/m_{i1}=8$, $n_{0i2}/n_{0i1}=0.01$, $\sqrt{m_{i1}V^2/k_BT_{i1}}=2.9$, $Z_1=Z_2=1$, $T_{i1}/T_e=0.1$, $T_{i2}/T_e=10$, and $\alpha=0.3$, which corresponds to the perturbations of plasma by a slow ion-acoustic wave. The Sagdeev pseudopotential $U_S(\varphi)$ shown by Fig.~\ref{fig:16+21} has two potential wells, that allows for the existence of small-amplitude linear, high-amplitude nonlinear, and large-amplitude (above the local maximum of $U_S(\varphi)$) super-nonlinear oscillations around the initial equilibrium $\varphi=0$. The oscillations of the plasma parameters, including the wave-induced fluxes $\Gamma_{1,2}(\xi)$ (\ref{eq:ei_flux}) of both sorts of ions oscillating in anti-phase, in these three regimes are shown in Fig.~\ref{fig:22+23+24}.
{Namely, the left-hand, middle, and right-hand columns of Fig.~\ref{fig:22+23+24} show oscillation profiles in the linear, nonlinear, and super-nonlinear regimes, respectively.}

\begin{figure}
	\begin{center}
		\includegraphics[width=0.8\linewidth]{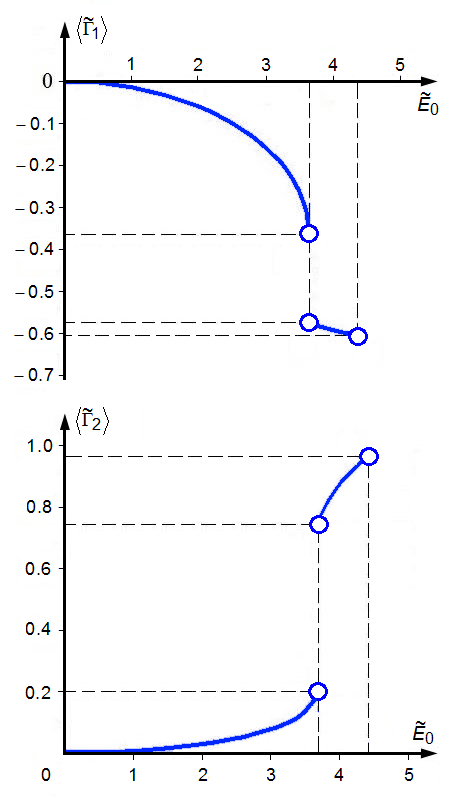}
	\end{center}
	\caption{The same as shown in Fig.~\ref{fig:6}, but for the slow ion-acoustic wave in the bi-ion \textit{eiid}-plasma. The unperturbed parameters of the ion component \lq\lq 1\rq\rq\ were used for normalisation.
	{The abrupt increase of the mean flux absolute value is caused by the transition to a super-nonlinear regime of the slow ion-acoustic wave (see Figs.~\ref{fig:16+21} and \ref{fig:22+23+24}).}
	}
	\label{fig:25}
\end{figure}

In Fig.~\ref{fig:25}, we show the dependence of the mean ion fluxes $\langle \Gamma_{1,2}\rangle$ (\ref{eq:ei_flux}) on the amplitude of the discussed slow ion-acoustic wave. It demonstrates that similarly to the pure (without dust) \textit{eii}-plasma considered in Sec.~\ref{sec:eii}, mean ion fluxes induced by a slow wave in the dusty \textit{eiid}-plasma are of different signs, i.e. counter-directed. Moreover, the absolute values of the mean ion fluxes are found to increase substantially (by $\sim$200--500\%) when the slow ion-acoustic wave switches to a super-nonlinear regime. 
{As seen from Figs.~\ref{fig:16+21} and \ref{fig:22+23+24}, the transition from a nonlinear regime to a super-nonlinear regime of ion-acoustic waves is characterised by a sudden increase in the oscillation amplitude and asymmetry with respect to the equilibrium value. This in turn causes the corresponding absolute value of the mean ion flux carried by a super-nonlinear ion-acoustic wave to increase abruptly as shown by Fig.~\ref{fig:25}.}

\section{Conclusions}\label{sec:conc}

We examined the ion fluxes induced by nonlinear ion-acoustic waves in the \textit{ei}-plasma with cold ions (Sagdeev's model), \textit{eii}-plasma with two types of warm positively charged ions, and dusty \textit{eiid}-plasma. 
The absolute values of the ion fluxes averaged over the wave period are shown to increase with the wave amplitude, while the flux sign, i.e. direction of the electrostatic wave-caused drift of ions, depends on the plasma composition and mode of the ion-acoustic wave.
{The latter, i.e. the discrimination between the fast and slow ion-acoustic modes in a multi-ion plasma, in turn depends on the combination of the parameters of the plasma and the wave. In this work, the combinations of the model parameters corresponding to the fast-mode and slow-mode regimes of the wave were chosen for illustration purposes only, with the main scope being to study the dependence of the revealed flux properties on the wave amplitude in each of those regimes.}
Thus, in the \textit{ei}-plasma there is a single ion-acoustic mode, in which the induced nonlinear ion flux is directed oppositely to the wave propagation. In contrast, the bi-ion \textit{eii}-plasma sustains two modes of ion-acoustic waves, fast and slow, in which ions of different types oscillate in-phase and in anti-phase, respectively.
In this case, the nonlinear fast ion-acoustic wave causes the fluxes of both types of ions which are co-directed and move against the wave. In the nonlinear slow ion-acoustic wave, the fluxes of ions are found to be of different signs, i.e. directed oppositely to each other, despite both of the ion populations were considered to have the same sign of the electric charge.

The revealed regime provides a unique opportunity to develop a new technology for separating gases by ion types or separating isotopes in plasma. We point out that the separation of non-volatile rare-earth isotopes is currently widely based on the method of ion-cyclotron resonance (ICR), in which one of the ion components is heated and then extracted from the plasma\cite{1976PhRvL..37.1547D, 2001PhRN..32.828D, 2009PhyU...52..345D}. In our case, the slow ion-acoustic wave will cause different types of ions to drift in different directions, which is technologically more advantageous than with the ICR separation method. In addition, the revealed ion-acoustic wave (IAW) separation method, in contrast to the ICR method, does not require sources of the external magnetic field. The proposed concept of the IAW separation can be readily adapted to plasma with a larger number of ion populations. For this, it is only necessary to correctly identify (according to Ref.~\onlinecite{2009PhyS...80c5504D}, for example) one of the many slow ion-acoustic wave modes, in which the targeted type of ions would oscillate in anti-phase with other types of ions.

We have also demonstrated that it is expedient to carry out the IAW-based separation of ions in a multicomponent plasma in a super-nonlinear regime of ion-acoustic waves. For example, the process of ion separation by a super-nonlinear slow ion-acoustic wave in the dusty \textit{eiid}-plasma  is shown to be several times more efficient in comparison with usual nonlinear waves in the pure \textit{eii}-plasma.

{The proposed theoretical model has a clear potential for a number of important developments and generalisations. For example, the model of a dusty plasma with a constant dust charge, considered in Sec.~\ref{sec:eiid}, could be extended upon the variable dust charge, treating the revealed nonlinear ion fluxes among the main charging processes of the dust grains (see e.g. Refs.~\onlinecite{2003NJPh....5...17S, 2008JTePh..53.1129D}). Another potentially interesting follow up work could be to perform a detailed analysis of the revealed ion fluxes and assess efficiency of the IAW-based separation of ions on the parameters of the plasma (e.g. the ion mass ratio) and the wave (e.g. the propagation speed).}

\begin{acknowledgments}
{The authors would like to dedicate this work in memory of Prof. George Rowlands (1932--2021) from the University of Warwick (UK), his inspiration and countless contribution to plasma physics and nonlinear wave dynamics.}
I.N.K. specially acknowledges support from the Foundation for the Advancement of Theoretical Physics and Mathematics “BASIS” (grant no. 19-1-5-58-1 for PhD students). D.Y.K. acknowledges support from the STFC consolidated grant ST/T000252/1 and the Ministry of Science and Higher Education of the Russian Federation.

\end{acknowledgments}

\section*{Data Availability Statement}
The data that supports the findings of this study are available within the article.


%

\appendix

\section{Pseudopotential $U_S(\varphi)$ for the \textit{eii}-plasma (Sec.~\ref{sec:eii})}
\label{app:eii_sagdeev}

{Integration of the electric charge density $\rho_{eii}(\varphi)$ (\ref{eq:eii_rho(phi)}) by $\varphi$ gives}

\begin{align}\label{eq:eii_sagdeev}
	&\frac{U_S(\varphi)}{4\pi k_B T_{i1}}=-\int_0^{\varphi}\frac{\rho_{eii}(\varphi)}{k_B T_{i1}}\,d\varphi =n_{0e}\frac{T_e}{T_{i1}}\left[1-\exp\left(-\frac{e\varphi}{k_BT_e}\right)\right]\nonumber\\
	&+n_{0i1}\frac{W_{0,-1}\left[-\dfrac{m_{i1}V^2}{k_BT_{i1}}\exp\left(-\dfrac{m_{i1}V^2}{k_B T_{i1}}\right)\right]+1}{\sqrt{-\dfrac{m_{i1}V^2}{k_BT_{i1}}W_{0,-1}\left[-\dfrac{m_{i1}V^2}{k_BT_{i1}}\exp\left(-\dfrac{m_{i1}V^2}{k_B T_{i1}}\right)\right]}}\\
	&-n_{0i1}\frac{W_{0,-1}\left[-\dfrac{m_{i1}V^2}{k_BT_{i1}}\exp\left(-\dfrac{m_{i1}V^2}{k_B T_{i1}} - \dfrac{2Z_1e\varphi}{k_BT_{i1}}\right)\right]+1}{\sqrt{-\dfrac{m_{i1}V^2}{k_BT_{i1}}W_{0,-1}\left[-\dfrac{m_{i1}V^2}{k_BT_{i1}}\exp\left(-\dfrac{m_{i1}V^2}{k_B T_{i1}} - \dfrac{2Z_1e\varphi}{k_BT_{i1}} \right)\right]}}\nonumber\\
	&+n_{0i2}\frac{\dfrac{T_{i2}}{T_{i1}}W_{0,-1}\left[-\dfrac{m_{i2}V^2}{k_BT_{i2}}\exp\left(-\dfrac{m_{i2}V^2}{k_B T_{i2}}\right)\right]+1}{\sqrt{-\dfrac{m_{i2}V^2}{k_BT_{i2}}W_{0,-1}\left[-\dfrac{m_{i2}V^2}{k_BT_{i2}}\exp\left(-\dfrac{m_{i2}V^2}{k_B T_{i2}}\right)\right]}}\nonumber\\
	&-n_{0i2}\frac{\dfrac{T_{i2}}{T_{i1}}W_{0,-1}\left[-\dfrac{m_{i2}V^2}{k_BT_{i2}}\exp\left(-\dfrac{m_{i2}V^2}{k_B T_{i2}} - \dfrac{2Z_2e\varphi}{k_BT_{i2}}\right)\right]+1}{\sqrt{-\dfrac{m_{i2}V^2}{k_BT_{i2}}W_{0,-1}\left[-\dfrac{m_{i2}V^2}{k_BT_{i2}}\exp\left(-\dfrac{m_{i2}V^2}{k_B T_{i2}} - \dfrac{2Z_2e\varphi}{k_BT_{i2}} \right)\right]}}\nonumber.
\end{align}

\section{Pseudopotential $U_S(\varphi)$ for the dusty \textit{eiid}-plasma (Sec.~\ref{sec:eiid})}
\label{app:eiid_sagdeev}

{Integration of the electric charge density $\rho_{eiid}(\varphi)$ (\ref{eq:eiid_rho(phi)}) by $\varphi$ gives}

\begin{align}\label{eq:eiid_sagdeev}
	&\frac{U_S(\varphi)}{4\pi k_B T_{i1}}=-\int_0^{\varphi}\frac{\rho_{eiid}(\varphi)}{k_B T_{i1}}\,d\varphi =n_{0e}\frac{T_e}{T_{i1}}\left[1-\exp\left(-\frac{e\varphi}{k_BT_e}\right)\right]\nonumber\\
	&+n_{0i1}\frac{W_{0,-1}\left[-\dfrac{m_{i1}V^2}{k_BT_{i1}}\exp\left(-\dfrac{m_{i1}V^2}{k_B T_{i1}}\right)\right]+1}{\sqrt{-\dfrac{m_{i1}V^2}{k_BT_{i1}}W_{0,-1}\left[-\dfrac{m_{i1}V^2}{k_BT_{i1}}\exp\left(-\dfrac{m_{i1}V^2}{k_B T_{i1}}\right)\right]}}\\
	&-n_{0i1}\frac{W_{0,-1}\left[-\dfrac{m_{i1}V^2}{k_BT_{i1}}\exp\left(-\dfrac{m_{i1}V^2}{k_B T_{i1}} - \dfrac{2Z_1e\varphi}{k_BT_{i1}}\right)\right]+1}{\sqrt{-\dfrac{m_{i1}V^2}{k_BT_{i1}}W_{0,-1}\left[-\dfrac{m_{i1}V^2}{k_BT_{i1}}\exp\left(-\dfrac{m_{i1}V^2}{k_B T_{i1}} - \dfrac{2Z_1e\varphi}{k_BT_{i1}} \right)\right]}}\nonumber\\
	&+n_{0i2}\frac{\dfrac{T_{i2}}{T_{i1}}W_{0,-1}\left[-\dfrac{m_{i2}V^2}{k_BT_{i2}}\exp\left(-\dfrac{m_{i2}V^2}{k_B T_{i2}}\right)\right]+1}{\sqrt{-\dfrac{m_{i2}V^2}{k_BT_{i2}}W_{0,-1}\left[-\dfrac{m_{i2}V^2}{k_BT_{i2}}\exp\left(-\dfrac{m_{i2}V^2}{k_B T_{i2}}\right)\right]}}\nonumber\\
	&-n_{0i2}\frac{\dfrac{T_{i2}}{T_{i1}}W_{0,-1}\left[-\dfrac{m_{i2}V^2}{k_BT_{i2}}\exp\left(-\dfrac{m_{i2}V^2}{k_B T_{i2}} - \dfrac{2Z_2e\varphi}{k_BT_{i2}}\right)\right]+1}{\sqrt{-\dfrac{m_{i2}V^2}{k_BT_{i2}}W_{0,-1}\left[-\dfrac{m_{i2}V^2}{k_BT_{i2}}\exp\left(-\dfrac{m_{i2}V^2}{k_B T_{i2}} - \dfrac{2Z_2e\varphi}{k_BT_{i2}} \right)\right]}}\nonumber\\
	&+\alpha n_{0i1}\frac{e\varphi}{k_B T_{i1}},\nonumber
\end{align}
{where the parameter $\alpha=q_dn_d/en_{0i1}$ describes the dusty fraction.}

\end{document}